\documentclass[twocolumn,showpacs,preprintnumbers,amsmath,amssymb]{revtex4}
\usepackage{amsmath,amsfonts,latexsym,amssymb,graphics,epsfig,subfigure,color,makeidx}

\begin{document}

\renewcommand{\baselinestretch}{1.3}

\title{Resonance spectrum of a bulk fermion on branes}

\author{Yu-Peng Zhang\footnote{zhangyupeng14@lzu.edu.cn},
        Yun-Zhi Du\footnote{duyzh13@lzu.edu.cn},
        Wen-Di Guo\footnote{guowd14@lzu.edu.cn},
        Yu-Xiao Liu \footnote{liuyx@lzu.edu.cn, corresponding author},
}
\affiliation{Institute of Theoretical Physics, Lanzhou University, Lanzhou 730000, China}

\begin{abstract}
 It is known that there are two mechanisms for localizing a bulk fermion on a brane, one is the well-known Yukawa coupling and the other is the new coupling proposed in [Phys. Rev. D 89, 086001 (2014)].  In this paper, we investigate localization and resonance spectrum of a bulk fermion on the same branes with the two localization mechanisms. It is found that both the two mechanisms can result in a volcano-like effective potential of the fermion Kaluza-Klein modes. The left-chiral fermion zero mode can be localized on the brane and there exist some discrete massive fermion Kaluza-Klein modes that quasilocalized on the brane (also called fermion resonances). The number of the fermion resonances increases linearly with the coupling parameter.
\end{abstract}

% \Keywords{ }


\maketitle

\section{Introduction}\label{scheme1}

It is well known that the Standard Model of particles and fields is not sufficient for interpreting some open questions such as the origin of the dark matter, the huge hierarchy between the weak and Planck scales, and the cosmological constant problem. Recently, there were many models that interpret the hierarchy problem \cite{Arkani-Hamed1998,Antoniadis1998,Randall1999,Gogberashvili2002,Das2008,Yang2012,Guo2015}, cosmological constant problem \cite{Arkani-Hamed2000,Starkman2001,JihnE.Kim2001,Gogberashvili2002a,Kehagias2004,Dey2009,Neupane2011}, and the dark matter \cite{Arkani-Hamed1999,Sahni2003,Cembranos2003} due to the existence of extra dimensions. The extra dimension theory was proposed in the 1920s by Kaluza and Klein (KK), who tried to unify the electromagnetism and Einstein's gravity by constructing a gravitational theory in a five-dimensional spacetime with a compact extra dimension \cite{Klein1926,KLEINOct91926,Gross1994}. Later, the KK theory was developed as braneworld theories \cite{Rubakov1983,Arkani-Hamed2000,Randall1999,Randall1999b}, where our universe is considered as a hypersurface (brane) embedded in the higher-dimensional spacetime.

In brane world scenarios, the particles and fields in the Standard Model should be confined on the brane. While the massive resonant KK modes (new particles beyond the Standard Model) can propagate into extra dimensions, which gives us the possibility of probing the extra dimensions through their interaction with particles on the brane \cite{Aaltonen2011,Aad2012,Sahin2015}.  The appearance of these resonances is related to the structure of the brane and the interaction between the bulk fields and the background fields generating the brane. Thus it is important and interesting to study the localization mechanisms and resonant KK modes on  thick branes with different internal structures. Usually, a thick brane can be generated by scalar fields  \cite{Gremm2000,Afonso2006,Bazeia2006,Bazeia2002,SouzaDutra2015} or vector fields \cite{Geng2015}. Localization and resonances of various bulk matter fields on a brane have been investigated in five-dimensional models \cite{Dvali1997,Randall1999,Randall1999b,Pomarol2000,Bajc2000,Oda2000,Gremm2000,Gregory2000,Ghoroku2002,Bazeia2004,Melfo2006,Liu2008,Liang2009,Guerrero2010,Castillo-Felisola2012,Jones2013,German2013,Andrianov2013,Sousa2013,Costa2013,Diaz-Furlong2014,Sousa2014,Rubin2015,Vaquera-2015,Choudhury2015,Jardim2015} and six-dimensional ones \cite{Parameswaran2007,Parameswaran2008,Gogberashvili2007,Costa2015,Arun2015,Dantas2015}.

Since the elementary matters consist of fermions, localization and resonances of spin 1/2 fermions are important in brane theories. In order to localize fermions on a thick brane of the RS-type, one usually needs to introduce some other interactions with background fields besides gravity. One simple interaction in a five-dimensional brane model is the Yukawa coupling between fermions and the background scalar fields \cite{Bajc2000,Ringeval2002,Melfo2006,Slatyer2007,Liu2008,Liu2009,Almeida2009,Liu2009a,Chumbes2011,Liu2011,Cruz2011,Correa2011,Castro2011,Castillo-Felisola2012,Andrianov2013,Barbosa2015,Agashe2015} when the scalar fields are odd functions of the extra dimension. With this coupling, the shapes of the effective potential of the left- or right-chiral fermion KK mode can be classified as three types: volcano-like \cite{Melfo2006,Ringeval2002,Almeida2009,Liu2009a,Liu2009,Cruz2011,Castro2011}, finite square well-like \cite{Liang2009,Zhao2010}, and harmonic potential-like \cite{Liu2008,Liu2011}. Correspondingly, the spectra of the KK fermions are continuous, partially discrete and partially continuous, and discrete. For the last two types of effective potentials, the fermion zero mode can be localized on the brane without further condition. While for the first one, the localization of the fermion zero mode usually needs that the Yukawa coupling is larger than some critical coupling ($\eta>\eta_{\text{c}}$). In all cases, the localized fermion zero mode is always chiral.

However, if the background scalar field is an even function of the extra dimension, the Yukawa coupling mechanism will do not work, since the $Z_2$ reflection symmetry of the effective potentials for the fermion KK modes can not be ensured \cite{Liu2014}. In this case, one should consider other mechanism. Recently, Liu, Xu, Chen, and Wei presented a new localization mechanism (shorted for the LXCW localization mechanism) for localizing bulk fermions on the brane generated by an even scalar field $\phi$ \cite{Liu2014}. The coupling is given by $\eta\bar{\Psi}\gamma_5\Gamma^M\partial_M{F(\phi)}\Psi$, which is used to describe the interaction between $\pi$-meson and nucleons in quantum field theory. The related study can be seen in Refs. \cite{Guo2015a,Xie2015}.
Interestingly, this new localization mechanism can also be used for the case of the brane generated by odd scalar fields.

Since there are two localization mechanisms for a bulk fermion, an interesting issue is whether the two mechanisms give similar results of the fermion localization. Therefore, the goal of this paper is to investigate localization of a bulk fermion on a same brane with the above two localization mechanisms that can yield the following results: (1) The fermion zero mode can be localized on the brane. (2) The effective potential of the left- or right-chiral fermion KK mode has the shape of volcano-like. (3) There exist fermion resonances that quasilocalized on the brane. To this end, we will consider two kinds of thick branes generated by one or at least one odd scalar field.

This paper is organized as follows. In Sec. \ref{review} we review the localization mechanisms and give the corresponding effective potentials of the fermion KK modes and the solutions of the fermion zero mode. In Sec. \ref{scheme2} we investigate localization and resonances of a bulk fermion on two kinds of thick branes: a single-scalar-field-generated thick brane and a multi-scalar-field-generated one. We make a simple comparison of results with the two different coupling mechanisms. Finally, we give a brief conclusion  in Sec. \ref{Conclusion}.

\section{Review of localization mechanism}{\label{review}}

In this section, we review the LXCW localization mechanism of a bulk fermion on a brane in five-dimensional spacetime, which is realized by introducing the coupling between the fermion field and the background scalar fields generating the brane. The corresponding  five-dimensional action is given by \cite{Liu2014}
    \begin{eqnarray}
    S_{\frac{1}{2}}=\int d^5x\sqrt{-g}~
       \Big[
           \bar{\Psi}\Gamma^M(\partial_M+\omega_M)\Psi\nonumber\\
           +\eta\bar{\Psi}\Gamma^M\partial_MF_1(\phi,\chi,\cdots,\rho)\gamma^5\Psi
       \Big],\label{action}
    \end{eqnarray}
where $F_1(\phi,\chi,\cdots,\rho)$ is a function of multiple scalar fields $\phi,~\chi,\cdots,~\rho$, and $\eta$ is the coupling constant. A five-dimensional Dirac fermion field is a four-component spinor and the corresponding gamma matrices $\Gamma^M$ satisfy $\{\Gamma^M,\Gamma^N\}=2g^{MN}$, where the five-dimensional spacetime indices are denoted by $M,~N,\cdots=0,~1,~2,~3,~5$. The spin connection $\omega_M$ is defined as
    \begin{equation}
    \omega_M=\frac{1}{4}\omega_M^{\,\,\,\,\bar{M}\bar{N}}\Gamma_{\bar{M}}\Gamma_{\bar{N}},
    \label{spin connection}
    \end{equation}
where
    \begin{eqnarray}
     \omega_M^{\,\,\,\,\bar{M}\bar{N}}
     &=&\frac{1}{2}E^{N\bar{M}}(\partial_ME^{\,\,\,\,\bar{N}}_N-\partial_NE^{\,\,\,\,\bar{N}}_M)\nonumber\\
     &&-\frac{1}{2}E^{N\bar{N}}(\partial_ME^{\,\,\,\,\bar{M}}_N-\partial_NE^{\,\,\,\,\bar{M}}_M)\nonumber\\
     &&-\frac{1}{2}E^{P\bar{M}}E^{Q\bar{N}}E^{\,\,\,\,\bar{R}}_M(\partial_P E_{Q\bar{R}}-\partial_Q E_{P\bar{R}}).\label{spin connection1}
    \end{eqnarray}
Here the letters $\bar{M},~\bar{N},\cdots$ are the five-dimensional local Lorentz indices and the vielbein $E^M_{\,\,\,\,\bar M}$ satisfies $E^M_{\,\,\,\,\bar{M}}E^N_{\,\,\,\,\bar{N}} \eta^{\bar{M}\bar{N}} = g^{MN}$. The relation between the gamma matrices ($\Gamma^M$) in a five-dimensional curved spacetime and the Minkowskian ones $\big(\Gamma^{\bar M}=(\Gamma^{\bar\mu},\Gamma^{\bar5})=(\gamma^{\bar\mu},\gamma^5)\big)$ is given by $\Gamma^M=E^M_{\,\,\,\,\bar M}\Gamma^{\bar M}$.

The metric of the five-dimensional spacetime describing a static braneworld system is given by \cite{Randall1999}
    \begin{eqnarray}
    ds^2&=&g_{MN}dx^Mdx^N \nonumber\\
        &=&e^{2A(y)}\hat{g}_{\mu\nu}(x^\lambda) dx^\mu dx^\nu+dy^2,\label{metric}
    \end{eqnarray}
where $y$ stands for the extra dimension, $e^{2A(y)}$ is the warp factor, and $\hat{g}_{\mu\nu}$ is the induced four-dimensional spacetime metric on the brane. Here, both the warp factor and the scalar fields are supposed to be functions of $y$ only, i. e. , $A=A(y),~ \phi=\phi(y),~\chi=\chi(y),~\cdots,~\rho=\rho(y)$. By performing the coordinate transformation $dy=e^A dz$, the line-element (\ref{metric}) can also be rewritten as
    \begin{equation}
    ds^2=e^{2A(z)}(\hat{g}_{\mu\nu}dx^\mu dx^\nu+dz^2),
    \label{cmetric}
    \end{equation}
which is a conformal flat metric if $\hat{g}_{\mu\nu}=\eta_{\mu\nu}$. For the brane metric  (\ref{cmetric}), non-vanishing components of the  spin connection (\ref{spin connection}) are  $\omega_\mu=\frac{1}{2}\partial_z A\gamma_\mu\gamma_5+\hat{\omega}_{\mu}$, where $\hat{\omega}_\mu$ is derived from the four-dimensional metric $\hat{g}_{\mu\nu}(x^\lambda)$.

From the conformal metric (\ref{cmetric}) and expression of the spin connection, the equation of  motion of the five-dimensional Dirac fermion $\Psi$ reads
    \begin{equation}
    \Big[\gamma^{\mu}(\partial_\mu+\hat{\omega}_{\mu})
    +\gamma^5(\partial_z+2\partial_z A)+\eta\partial_z F_1\Big]\Psi=0,
    \label{newcouplingmotion}
    \end{equation}
where $\gamma^{\mu}(\partial_\mu+\hat{\omega}_{\mu})$ is the dirac operator on the brane. While for the Yukawa coupling $-\eta \bar{\Psi}F_2(\phi,\chi,\cdots)\Psi$, the corresponding Dirac equation is given by \cite{Liu2008}
    \begin{equation}
    \Big[\gamma^{\mu}(\partial_\mu+\hat{\omega}_{\mu})
    +\gamma^5(\partial_z+2\partial_z A)+\eta e^{A}F_2\Big]\Psi=0.
    \label{yukawacouplingmotion}
    \end{equation}
Next, we make the following chiral decomposition for the five-dimensional Dirac field $\Psi$
    \begin{equation}
    \Psi(x,z)=e^{-2A}\sum_{n}\Big[\psi_{Ln}(x)f_{Ln}(z)
                       +\psi_{Rn}(x)f_{Rn}(z)\Big],
    \label{decomposition}
    \end{equation}
where $\psi_{Ln}=-\gamma^5\psi_{Ln}$ and $\psi_{Rn}=\gamma^5\psi_{Rn}$ are the left- and right-chiral components of the four-dimensional effective Dirac fermion field, respectively. Substituting the chiral decomposition (\ref{decomposition}) into Eq. (\ref{newcouplingmotion}), we get the four-dimensional massive Dirac equations
    \begin{eqnarray}
    \begin{array}{c}
      \gamma^{\mu}(\partial_\mu+\hat{\omega}_{\mu})\psi_{Ln}(x)=\mu_n\psi_{Rn}(x),\\
      \gamma^{\mu}(\partial_\mu+\hat{\omega}_{\mu})\psi_{Rn}(x)=\mu_n\psi_{Ln}(x),
    \end{array}
    \end{eqnarray}
and the following coupling equations of the KK modes $f_{Ln,Rn}$
    \begin{eqnarray}
    \begin{array}{c}
      (\partial_z -\eta\partial_z F)f_{Ln}=+\mu_nf_{Rn},\\
      (\partial_z +\eta\partial_z F)f_{Rn}=-\mu_nf_{Ln},
    \end{array}
    \label{couplefunction}
    \end{eqnarray}
where $\mu_n$ is the mass of the four-dimensional fermion fields $\psi_{Ln,Rn}(x)$. The above two equations can also be rewritten as
the Schr\"{o}dinger-like equations
    \begin{eqnarray}
    [-\partial_z^2+V_L(z)]f_{Ln} &=&\mu^2_nf_{Ln}, \label{schrodingerlikeequationl} \\  ~
    [-\partial_z^2+V_R(z)]f_{Rn} &=&\mu^2_nf_{Rn}
    \label{schrodingerlikeequationr}
    \end{eqnarray}
with the effective potentials
    \begin{equation}
    V_{L,R}(z)=(\eta\partial_zF_1)^2\pm\partial_z(\eta\partial_zF_1).
    \label{potentialnew}
    \end{equation}
Note that the Schr\"{o}dinger-like equations (\ref{schrodingerlikeequationl}) and (\ref{schrodingerlikeequationr}) can be transformed as
    \begin{eqnarray}
     \begin{array}{c}
       U^{\dag}U f_{Ln}=\mu^2_nf_{Ln} \\
       UU^{\dag} f_{Rn}=\mu^2_nf_{Rn}
     \end{array}
    \label{operator2}
    \end{eqnarray}
with the operator $U=\partial_z-\eta\partial_z F$, which insure that the mass square is non-negative, i.e., $\mu_n^2 \ge 0$.

For the Yukawa coupling $-\eta F_2(\phi,\chi,\cdots)\bar{\Psi}\Psi$, the corresponding effective potentials in the Schr\"{o}dinger-like equations are \cite{Liu2008}
    \begin{equation}
    V_{L,R}(z)=(\eta e^AF_2)^2\pm\partial_z(\eta e^AF_2).
    \label{potentialyukawa}
    \end{equation}

In order to obtain the four-dimensional effective action for the massless and massive Dirac fermions
    \begin{eqnarray}
    S_{\text{eff}}=\sum_n\int d^4x\sqrt{-\hat g}~
           {\bar\psi_n}\big[
                \gamma^\mu(\partial_\mu+\omega_\mu)
               -\mu_n
           \big] \psi_n,
    \end{eqnarray}
the KK modes $f_{Ln,Rn}$ should satisfy the orthonormality conditions
    \begin{eqnarray}
    \int_{-\infty}^{+\infty} {f_{Lm}f_{Ln}dz}
      &=&\int_{-\infty}^{+\infty} {f_{Rm}f_{Rn}dz}
      =\delta_{mn},\nonumber\\
    \int_{-\infty}^{+\infty} {f_{Ln}f_{Rn}dz}&=&0,\label{orthonormality}
    \end{eqnarray}
which are important to check whether the KK modes can be localized on the brane. From Eq. (\ref{couplefunction}), the corresponding chiral zero-modes read
    \begin{equation}
    f_{L0,R0}\propto
    e^{ \pm\eta\int {dz}~\partial_z F_1 }
    =e^{\pm\eta F_1}. \label{newzero mode}
    \end{equation}
While for the Yukawa coupling mechanism, the zero-modes are
    \begin{equation}
    f_{L0,R0} \propto
    e^{\pm\eta \int {dz}~ e^{A}   F_2(z)}
    =e^{\pm\eta\int {dy} ~F_2(y)}.\label{yukawazero mode}
    \end{equation}
In the following we will use the subscripts ``new" and ``Yuk" for the LXCW and Yukawa coupling mechanisms, respectively.

The massive KK modes can be obtained by solving Eqs. (\ref{schrodingerlikeequationl}) numerically. Inspired by the results of Ref. \cite{Liu2008}, Almeida \emph{ea al} investigated the issue of localization
of a bulk fermion on a brane and firstly suggested that large peaks in the distribution of the normalized squared wavefunction $|f_{L,R}(0)|^2$ as a function of $m$ would reveal the existence of fermion resonant states \cite{Almeida2009}. However, this method is effective only for even fermion resonances because $f_{L,R}(0)=0$ for any odd wavefunction. In order to find all fermion resonances, Liu \emph{et al} presented a new method called relative probability, which is defined as \cite{Liu2009a}:
    \begin{eqnarray}
    P=\frac{\int_{-z_b}^{z_b}|f_{Ln,Rn}(z)|^2 dz}
      {\int_{z_{max}}^{-z_{max}}|f_{Ln,Rn}(z)|^2 dz},
      \label{zbmin}
    \end{eqnarray}
where $z_{max}=10z_b$ and the parameter $z_b$ could be chosen as the coordinate of the maximum of the corresponding effective potential. Since the potentials considered in this paper are symmetric, the wave functions are either even or odd. Hence, we can use the following boundary conditions to solve the differential equation  (\ref{schrodingerlikeequationl}) numerically~\cite{Liu2009a}:
    \begin{eqnarray}
    \label{incondition}
      f_{Ln,Rn}(0)&=&0, ~f'_{Ln,Rn}(0)=1,~\text{odd KK modes},\\
      f_{Ln,Rn}(0)&=&1, ~f'_{Ln,Rn}(0)=0,~\text{even KK modes}.
    \end{eqnarray}

Next we will investigate localization and resonances of the fermions in two thick braneworld models.

\section{Fermion localization and resonances}\label{scheme2}

\subsection{Single-scalar-field thick brane}

In this subsection, we investigate localization and resonances of a bulk fermion on a single-scalar-field thick brane in a five-dimensional spacetime \cite{Afonso2006,Liang2009}. The action of this system reads
    \begin{eqnarray}
    S=\int{d^5x\sqrt{-g}~\left[\frac{M_5^3}{4}R-\frac{1}{2}\partial_M\phi\partial^M\phi-V(\phi)\right]},
    \end{eqnarray}
where $R$ is the five-dimensional scalar curvature and the fundamental mass scale $M_5$ will be set to 1 for convenience.

The line element (\ref{metric}) can be rewritten as
    \begin{eqnarray}
    ds_5^2=e^{2A}ds_4^2+dy^2,\label{Is}
    \end{eqnarray}
where $ds_4^2$ stands for the line element on the brane. For the warped thick branes, $ds_4^2$ has the following forms
    \begin{eqnarray}
    d s_4^2=\left\{
    \begin{array}{cc}
    -dt^2+e^{2\beta t}(dx_1^2+dx_2^2+dx_3^2)~~~\textrm{$\text{dS}_4$ brane},\label{ds}\\
    e^{-2\beta x_3}(-dt^2+dx_1^2+dx_2^2)+dx_3^2~~~\textrm{$\text{AdS}_4$ brane}
    \label{Ads}.
    \end{array}\right.
    \end{eqnarray}
Here $\beta$ is a parameter related to the four-dimensional cosmological constant of the dS$_4$ or AdS$_4$ brane: $\Lambda_4=3\beta^2$ or $\Lambda_4=-3\beta^2$ \cite{Liu2010,Liu2011}.

A brane solution was found in Ref. \cite{Afonso2006}:
    \begin{eqnarray}
    V(\phi)&=&\frac{3}{4}a^2(1+\Lambda_{4}) \big[1+(1+3s)\Lambda_{4} \big]\cosh^2(b\phi)\nonumber\\
             &-&3a^2(1+\Lambda_{4})^2\sinh^2(b\phi),\label{phi}\\
    \phi(y)&=&\frac{1}{b}\text{arcsinh}(\tan \bar{y} ),\label{potentialphi}\\
    A(y)&=&-\frac{1}{2}\ln[sa^2(1+\Lambda_{4})\sec^2 \bar{y} ],\label{warpfactorscalar}
    \end{eqnarray}
where  $\bar{y}\equiv a(1+\Lambda_{4})y$, the parameter $s\in(0,1]$, $a$ is a real parameter, and $b=\sqrt{\frac{2(1+\Lambda_{4})}{3(1+(1+s)\Lambda_{4})}}$. It is obvious to see that the thick brane is extended in the interval $y\in\left(-|\frac{\pi}{2a(1+\Lambda_{4})}|,~|\frac{\pi}{2a(1+\Lambda_{4})}|\right)$. The relation between the conformal and physical coordinates reads
    \begin{eqnarray}
    z=\int_0^{y}e^{-A(\hat{y})} d\hat{y}
     =\frac{1}{h} \ln\left(\frac{\cos\big(\frac{1}{2}  \bar{y}\big)+\sin\big(\frac{1}{2} \bar{y}\big)}{\cos\big(\frac{1}{2} \bar{y}\big)-\sin\big(\frac{1}{2}  \bar{y}\big)}\right), \label{zy}
    \end{eqnarray}
from which we have
    \begin{eqnarray}
    y=\frac{1}{a(1+\Lambda_{4} )}
       \left[2\arctan\left(e^{hz}\right)
             -\frac{\pi }{2}
       \right].              \label{yz}
    \end{eqnarray}
where $h\equiv\sqrt{\frac{1+\Lambda_{4} }{s}}$.
Note that the conformal coordinate $z$ will trend to infinite as $y\rightarrow|\frac{\pi}{2a(1+\Lambda_{4})}|$.

We can substitute the relation (\ref{yz}) into Eqs.~(\ref{warpfactorscalar}) and (\ref{phi}) to obtain the corresponding warp factor and scalar field in the coordinate $z$ \cite{Liang2009}:
    \begin{eqnarray}
    \phi(z)&=&\frac{1}{b}\text{arcsinh} \left[\sinh(hz)\right]
        =\frac{h}{b}z,\nonumber\\
    A(z)&=&-\frac{1}{2} \ln\big[a^2 s (1+\Lambda_{4} ) \cosh^2(hz)\big].\label{warpfactorscalarz}
    \end{eqnarray}
Note that $\Lambda_{4}=0$ corresponds to the flat brane case.

\subsubsection{\textbf{Yukawa coupling mechanism}}

\begin{figure}[!htb]
    \includegraphics[width=0.22\textwidth]{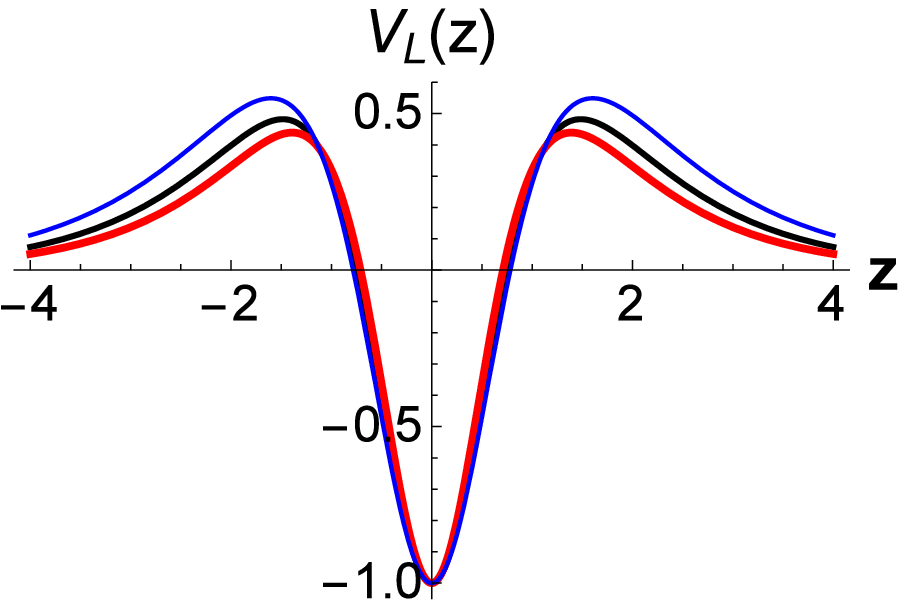}
    \includegraphics[width=0.22\textwidth]{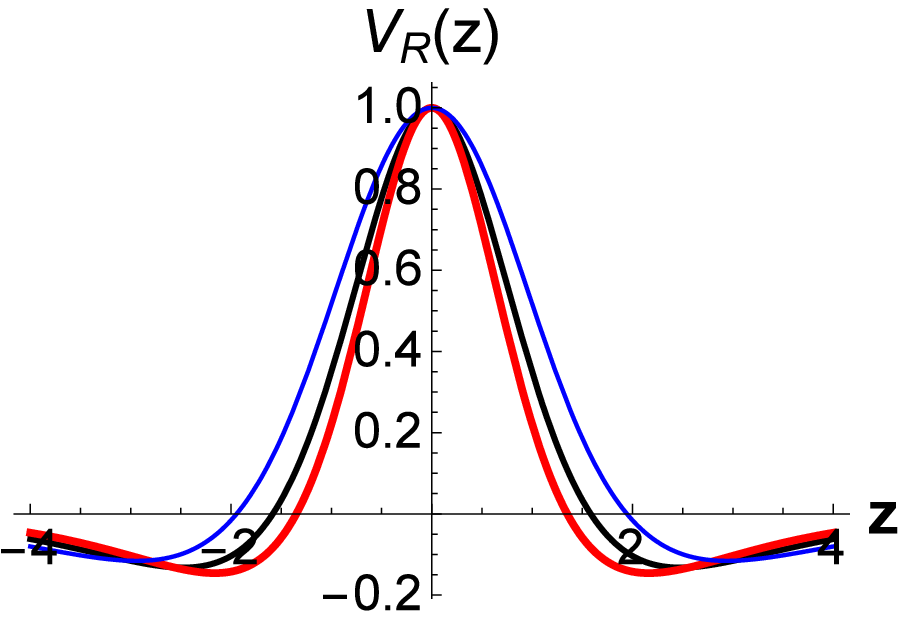}
    \vskip -4mm \caption{Plots of the effective potentials (\ref{potential_q}) with the Yukawa coupling ($F_2=\text{arcsinh}^{2q-1}(b\phi)$) for different values of $\Lambda_{4}$. The parameters are set to $a=s=q=-\eta=1$ and $\Lambda_{4}=0.2$~(thick red line), $0$~(black line), $-0.2$~(thin blue line). }
    \label{FigPotenialsYukLambda}
    \end{figure}

\begin{figure}[!htb]
    \includegraphics[width=0.22\textwidth]{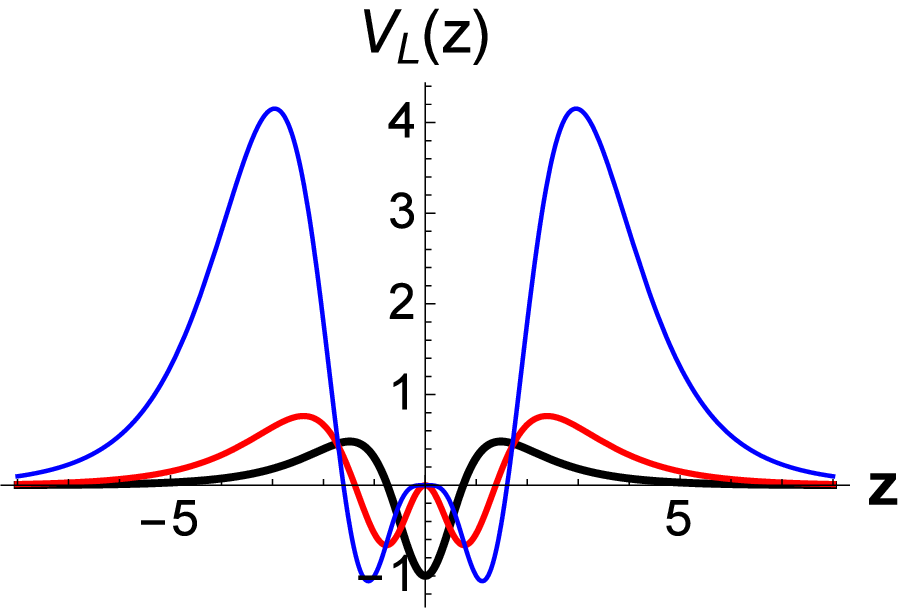}
    \includegraphics[width=0.22\textwidth]{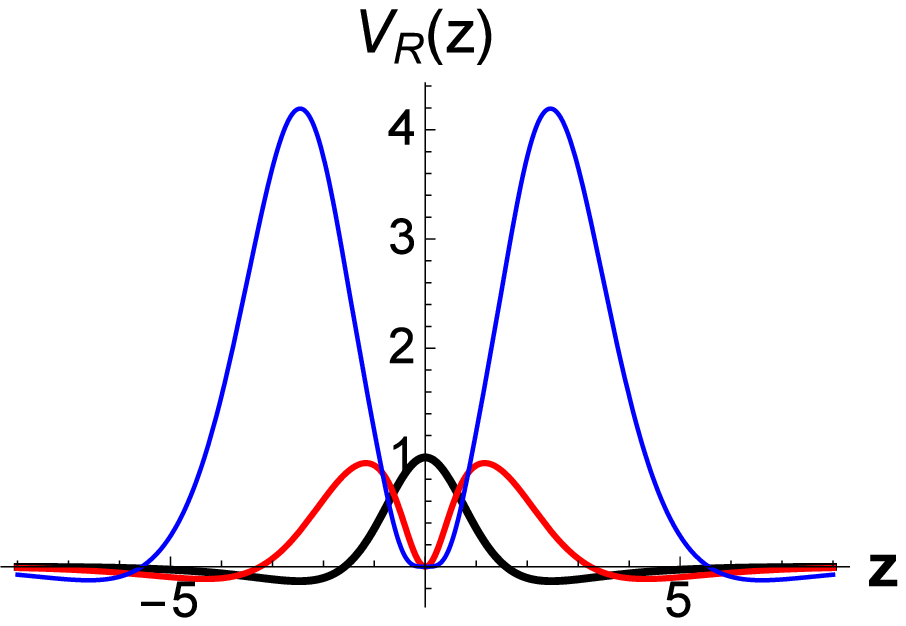}
    \vskip -4mm \caption{Plots of the effective potentials (\ref{potential_q}) with the Yukawa coupling ($F_2=\text{arcsinh}^{2q-1}(b\phi)$) for different values of $q$. The parameters are set to $a=s=q=-\eta=1$, $\Lambda_{4}=0$, and $q=1$~(thick black line), $2$~(red line), $3$~(thin blue line).}
    \label{FigPotenialsYuk_q}
    \end{figure}

Firstly, we consider the Yukawa coupling mechanism for the fermion localization on the single-scalar-field thick brane. The authors of Ref.~\cite{Liang2009} investigated localization of fermion on the brane with the forms of $F_2=\phi$ and $F_2=\sinh(b\phi)$, for which the effective potentials are respectively volcano-type and PT-type, but they did not consider fermion resonances. Here, we would like to investigate fermion resonances on the brane. The coupling function can be chosen as $F_2=\lambda_1 \phi + \lambda_2\sinh(b\phi) + \lambda_3 \text{arcsinh}^{2q-1}(b\phi)$ with a structure parameter $q$ (positive integer). For simplicity, we only consider the case of $\lambda_1=\lambda_2=0$, $\lambda_3=1$. From Eqs. (\ref{potentialyukawa}) and (\ref{yukawazero mode}), the effective potentials and zero modes are given by
    \begin{eqnarray}
    V_{L,R}^{\text{Yuk}}(z)&=&\pm\eta~\frac{\text{sech}\left(h z \right)\text{arcsinh}^{2q-2}
       \left(hz \right)}{\sqrt{a^2 s (1+\Lambda_{4} )}}
     \Bigg(\frac{(2q-1) h}
                {b \sqrt{1+h^2z^2} }  \nonumber\\
     &&\pm\frac{\eta~\text{arcsinh}^{2 q}\left( h z\right) \text{sech}\left(h z \right)}
               {\sqrt{a^2 s (1+\Lambda_{4} )}}\nonumber\\
     && -h~\text{arcsinh}\left(hz\right) \tanh\left(h z\right)
    \Bigg),    \label{potential_q}
    \end{eqnarray}
and
    \begin{equation}
    f_{L0,R0}^{\text{Yuk}}(z)\propto \exp\left[\mp\eta  \int \frac{\text{arcsinh}^{2q-1}\left(h{z}\right)}{\sqrt{a^2 s (1+\Lambda_{4} ) }\cosh\left(h{z} \right)} {dz}\right].
    \label{zero modenew}
    \end{equation}
Note that the parameters $\eta$, $q$, $a$ and $\Lambda_{4}$ affect the height and width of the effective potentials. Plots of the effective potentials for different values of $\Lambda_{4}$ and $q$ are shown in Figs.~\ref{FigPotenialsYukLambda} and \ref{FigPotenialsYuk_q}, respectively.

In this subsection we mainly consider localization and resonances of a bulk fermion on the flat brane ($\Lambda_{4}=0$) with different values of the positive integer $q$ (corresponds to different coupling functions). For simplicity, we choose $a=s=1$, for which $\bar{y}\equiv ay(1+\Lambda_{4})=y$. For $q=1$, the effective potentials have the form
    \begin{eqnarray}
    V_{L,R}^{\text{Yuk}}(z)&=&\pm~\eta~\text{sech}^2{z}\bigg(\frac{1}{\sqrt{1+z^2}}\pm\eta ~\text{arcsinh}^2{z}\nonumber\\
    &-&\text{arcsinh}{z} \sinh{z}\bigg).
    \label{potential_q=1}
    \end{eqnarray}
The behaviors of the effective potentials (\ref{potential_q=1}) for the Yukawa coupling mechanism are
    \begin{eqnarray}
    V_{L,R}^{\text{Yuk}}(0)&=&\pm\eta,\\
    V_{L,R}^{\text{Yuk}}(z\to\infty) &\to& \mp 4~\eta~e^{-z} \ln (2z) \to 0^{\pm}, \label{Vinfinite}
    \end{eqnarray}
where we have considered $\eta<0$ in the formula (\ref{Vinfinite}) in order to localize the left-chiral fermion zero mode.
The form of $F_2=\text{arcsinh}^{2q-1}(b\phi)$ is
    \begin{eqnarray}
    F_2(y)&=&\text{arcsinh} z.
    \end{eqnarray}
The zero modes read
    \begin{eqnarray}
    f_{L0,R0}^{\text{Yuk}}(z)& \propto  &\exp\left[\pm\eta\int \frac{\text{arcsinh}
    \left({z}\right)}{\text{cosh}({z})}d{z}\right].
    \end{eqnarray}
In order to obtain the fermion zero modes on the brane, the following normalization conditions should be satisfied
    \begin{eqnarray}
    \int_{-\infty}^\infty{|f_{L0,R0}|^2dz}<\infty.
    \label{zero modeyukawa}
    \end{eqnarray}
Because the integral $\int \frac{\text{arcsinh}\left({z}\right)}{\text{cosh}({z})}d{z} \to \int 2\ln(2z) e^{-z}d{z} = 2 \text{Ei}(-z) - 2e^{-z} \ln(2 z)\to 0$, where the function $\text{Ei}(z)=-\int_{-z}^{\infty } \left.e^{-t}\right/t \, dt$ and $\text{Ei}(-\infty)=0$, the zero modes trend to a constant at infinite. So the above normalization conditions can not be satisfied for both the two zero modes, which means that the zero modes can not be localized on the brane for the case of $\lambda_1=\lambda_2=0$.
So, in order to localize the left- or right-chiral fermion zero mode on the brane, we need to keep $\lambda_1$ or $\lambda_2$, or both or them. However, since our interesting is fermion resonances, we can neglect them for simplicity. Surely, if consider nonvanishing $\lambda_1$ or $\lambda_2$, the following numerical results will be different, but our conclusion will not change.

\begin{figure}[!htb]
    \includegraphics[width=0.22\textwidth]{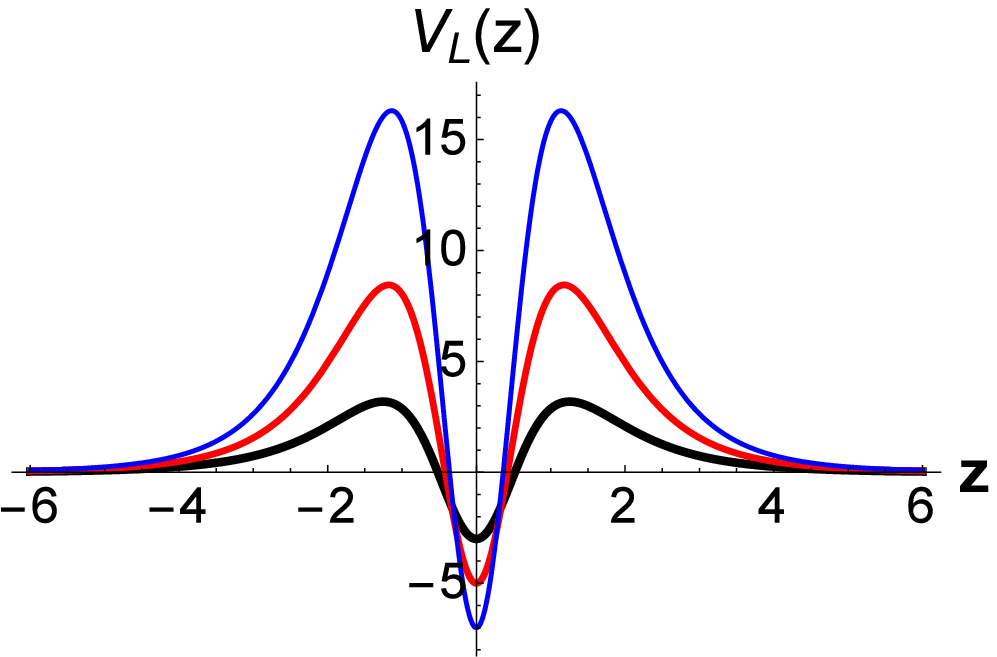}
    \includegraphics[width=0.22\textwidth]{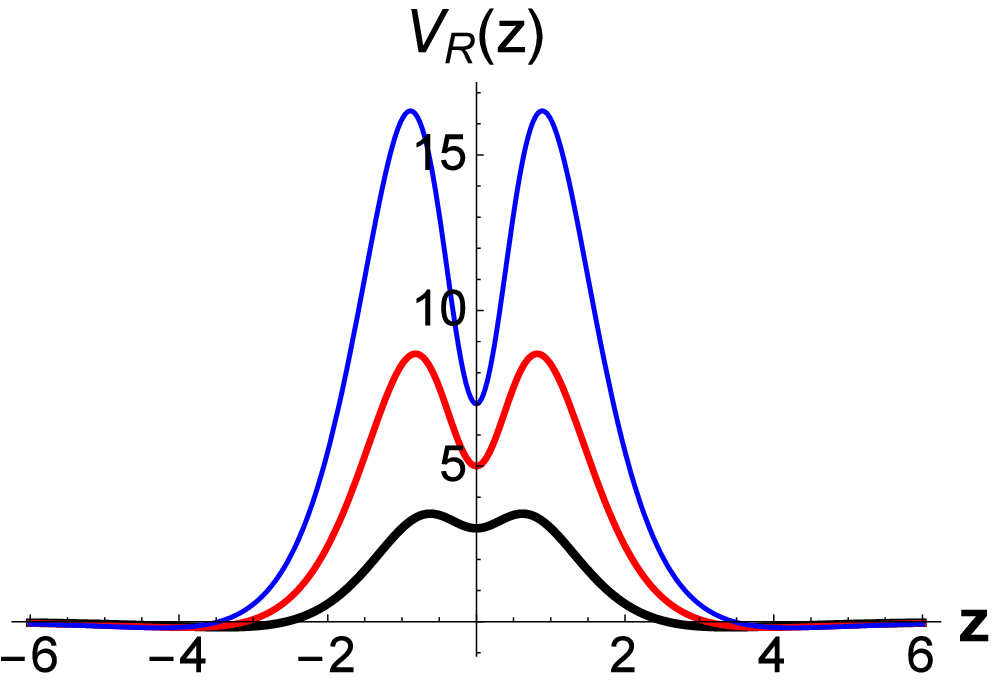}
    \vskip -4mm \caption{Plots of the effective potentials $V_{L,R}(z)$ (\ref{potential_q}) with the Yukawa coupling  for $q=1$ and different values of $\eta$. The parameters are set to $a=s=1$, $\Lambda_{4}=0$, and $\eta=-3$~(thick black line),~$-5$ (red line),~$-7$ (thin blue line). }
    \label{figneweq1}
    \end{figure}

\begin{figure}[!htb]
    \includegraphics[width=0.22\textwidth]{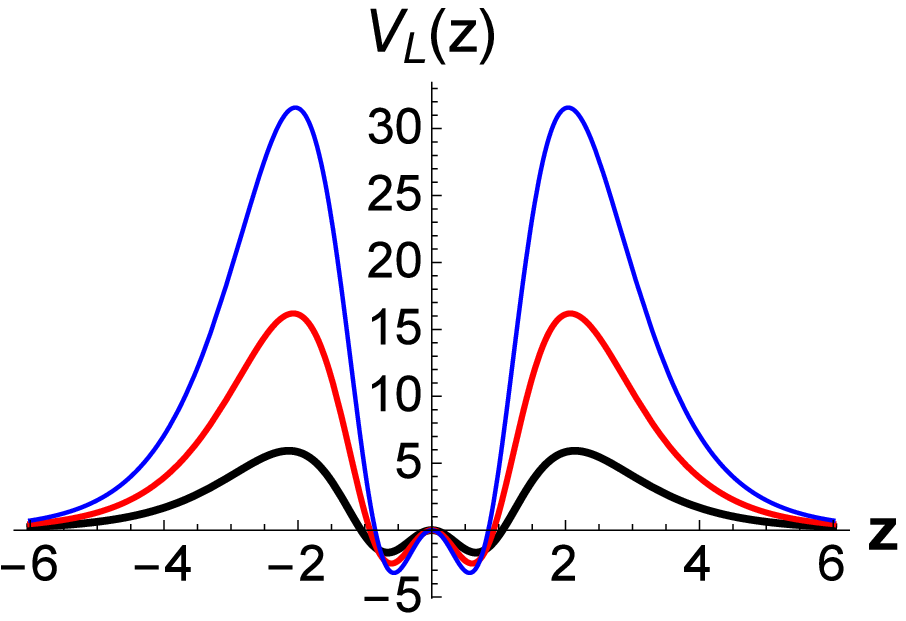}
    \includegraphics[width=0.22\textwidth]{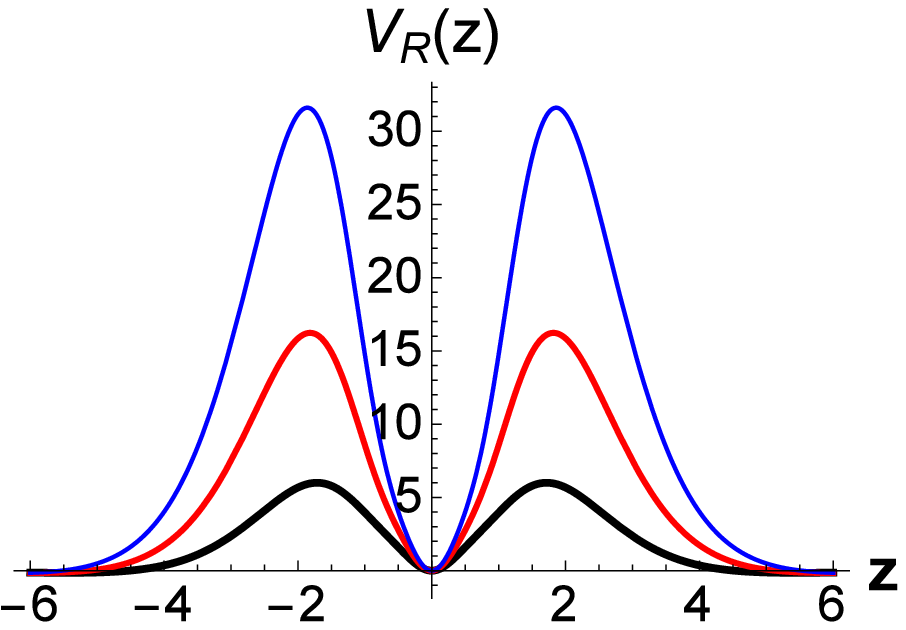}
    \vskip -4mm \caption{Plots of the effective potentials (\ref{potential_q}) with the Yukawa coupling  for  $q=2$ and  different values of $\eta$. The parameters are set to $a=s=1,~\Lambda_{4}=0$, and $\eta=-3$~(thick black line),~$-5$ (red line), $-7$~(thin blue line).}
    \label{figneweq2}
    \end{figure}

We can obtain the solution of the massive fermion KK modes by solving numerically the Schr\"{o}dinger-like equations (\ref{schrodingerlikeequationl}) and (\ref{schrodingerlikeequationr}) with the potentials (\ref{potential_q}). First, we choose $q=1,~2$ and $\eta=-3,~-5,~-7$, and the corresponding plots of the potentials are shown in Figs. \ref{figneweq1} and \ref{figneweq2}. From these figures, we can see that the depth of the quasi-well of the effective potential $V_R(z)$ increases with the parameters $|\eta|$ and $q$. This appearance of the quasi-well would result in quasi-localized massive KK fermions (also called fermions resonances).

\begin{figure}[!htb]
    \subfigure[$\eta=-3$]{
    \includegraphics[width=0.22\textwidth]{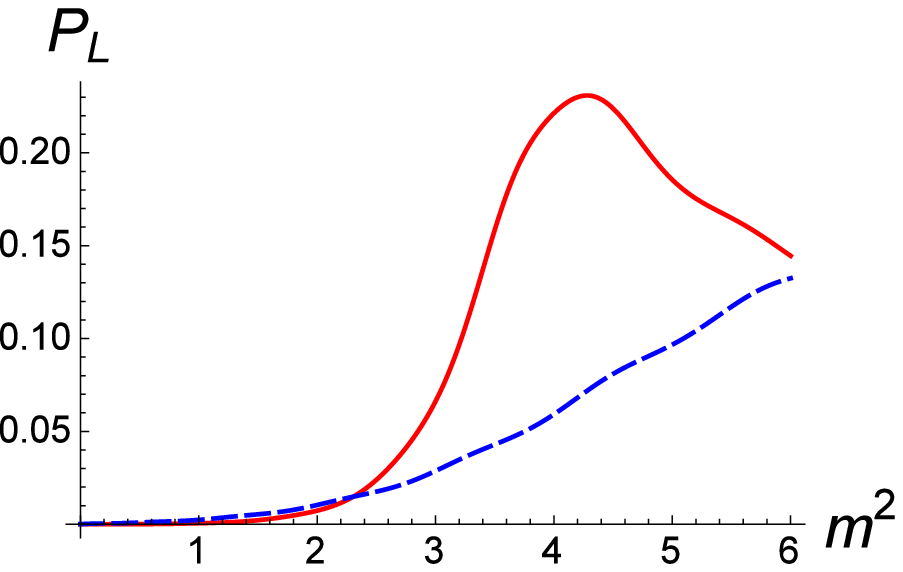}}
    \subfigure[$\eta=-3$]{
    \includegraphics[width=0.22\textwidth]{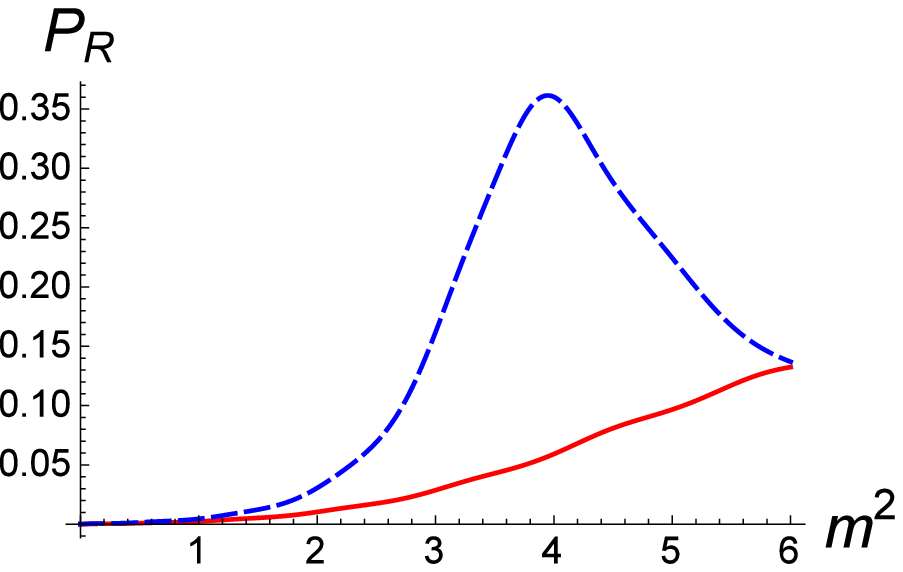}}
    \subfigure[$\eta=-5$]{
    \includegraphics[width=0.22\textwidth]{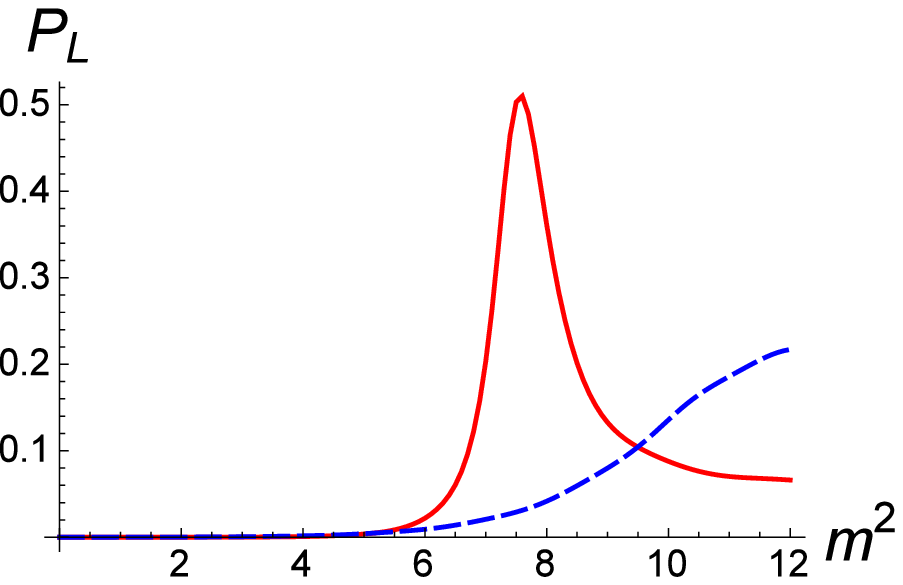}}
    \subfigure[$\eta=-5$]{
    \includegraphics[width=0.22\textwidth]{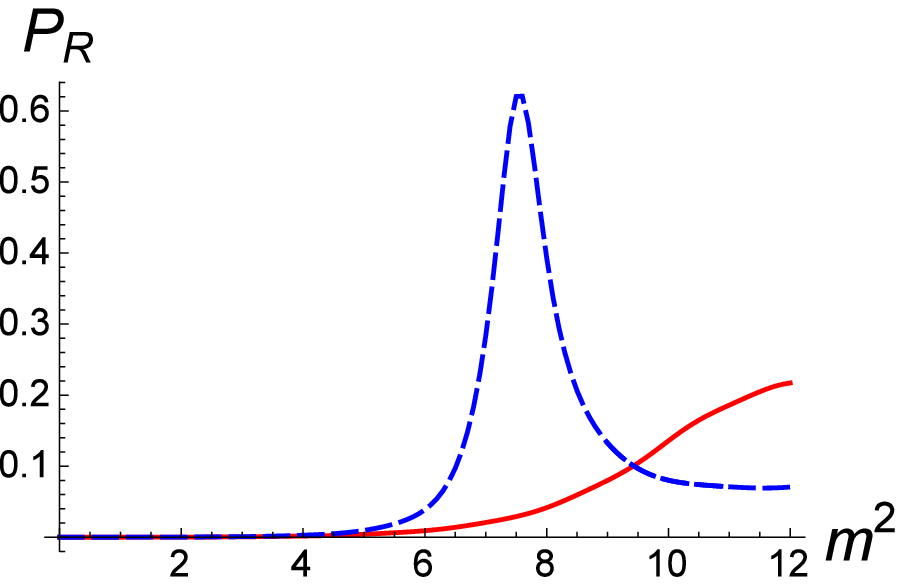}}
    \subfigure[$\eta=-7$]{
    \includegraphics[width=0.22\textwidth]{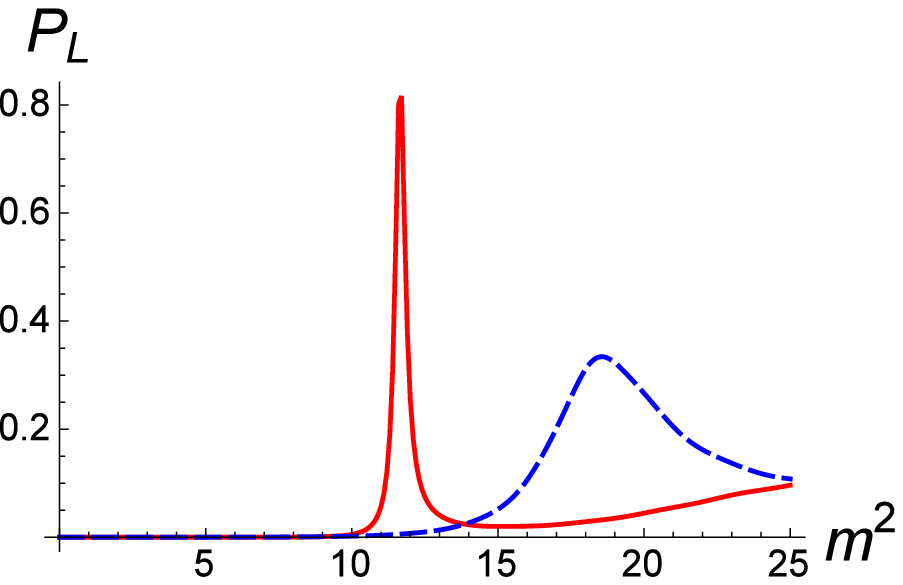}}
    \subfigure[$\eta=-7$]{
    \includegraphics[width=0.22\textwidth]{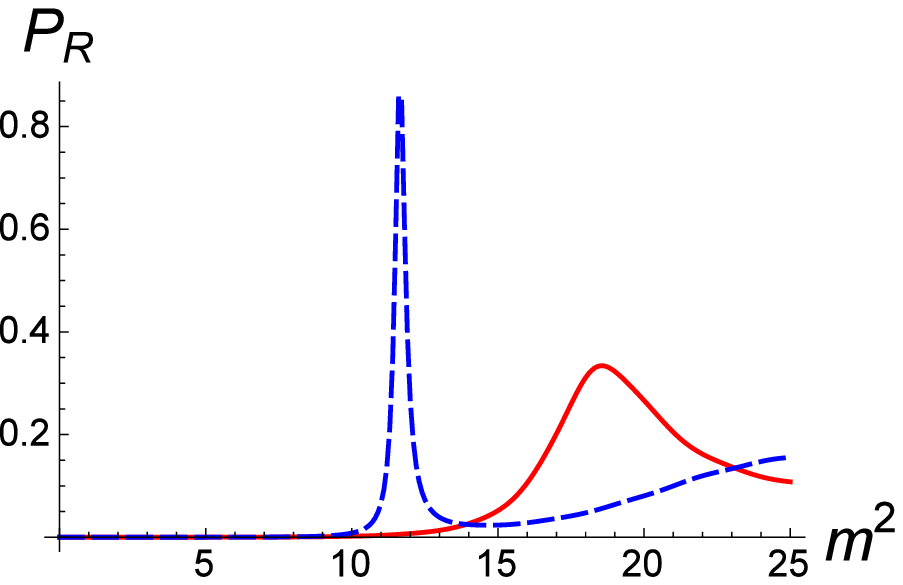}}
    \vskip -4mm \caption{Plots of the probabilities $P_{L,R}$ (as a function of $m^2$) for the Yukawa coupling mechanism, which include the even parity (blue dashed line) and odd parity (red line) massive KK modes of the left-chiral (up channel) and right-chiral (down channel) fermions. The parameters are set to $a=s=1$, $q=1$, and $\Lambda_{4}=0$.}
    \label{q1peaksl2}
    \end{figure}

\begin{figure}[!htb]
    \subfigure[$f_{L}(z)$]{
    \includegraphics[width=0.22\textwidth]{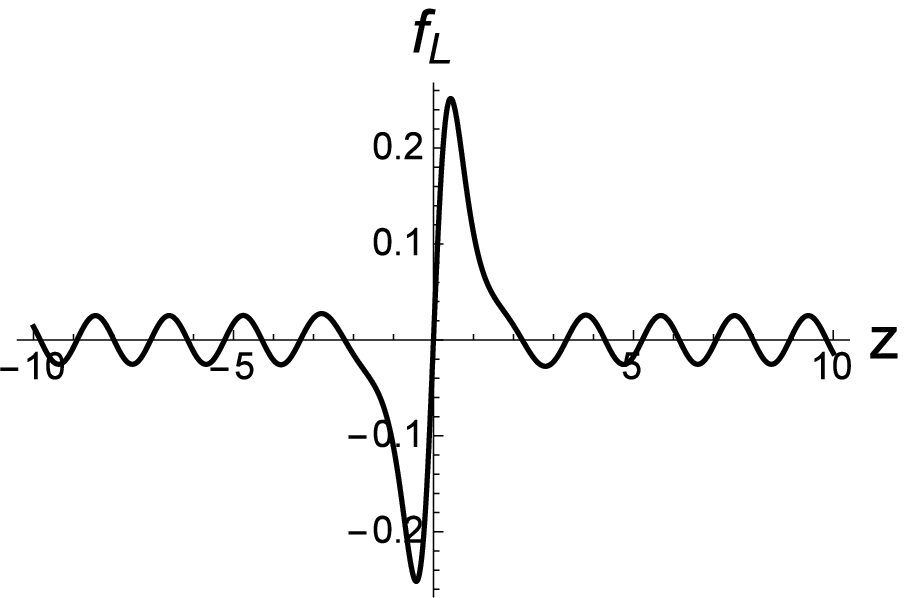}}
    \subfigure[$f_{R}(z)$]{
    \includegraphics[width=0.22\textwidth]{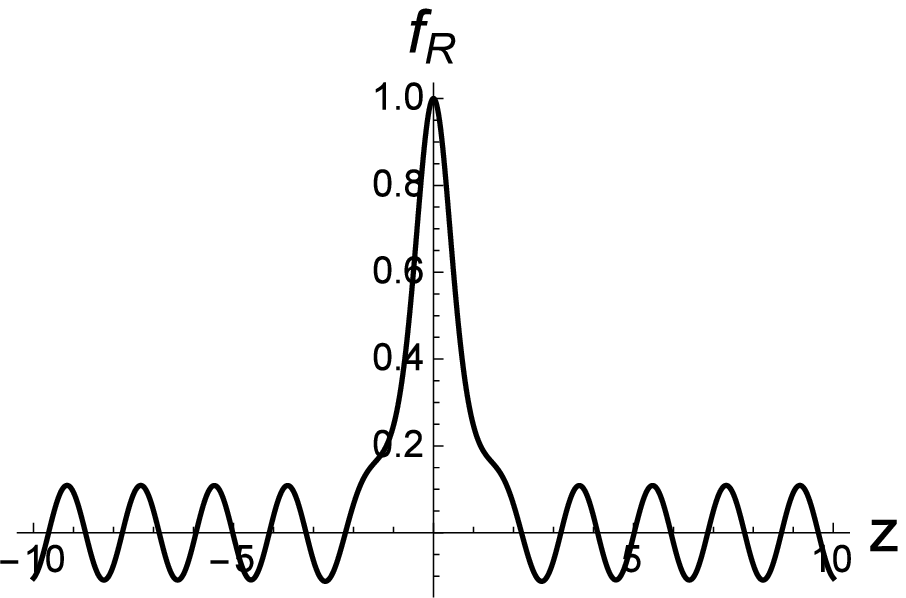}}
    \vskip -4mm \caption{Plots of the resonances of the left- and right-chiral KK fermions for the Yukawa coupling mechanism with $\eta=-7$ and $q=1$.}
    \label{ryesonances}
    \end{figure}

\begin{figure}[!htb]
    \subfigure[$\eta=-3$]{
    \includegraphics[width=0.22\textwidth]{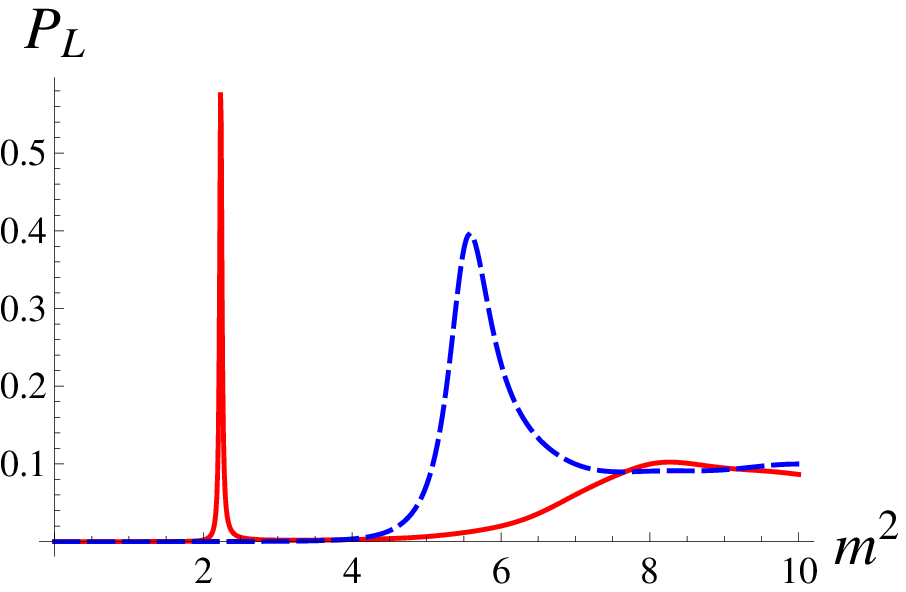}}
    \subfigure[$\eta=-3$]{
    \includegraphics[width=0.22\textwidth]{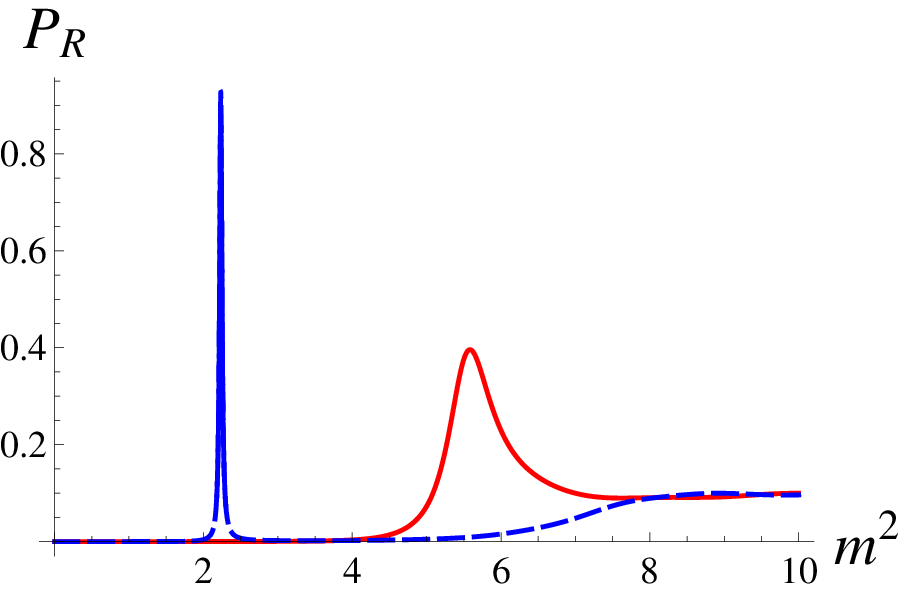}}
    \subfigure[$\eta=-5$]{
    \includegraphics[width=0.22\textwidth]{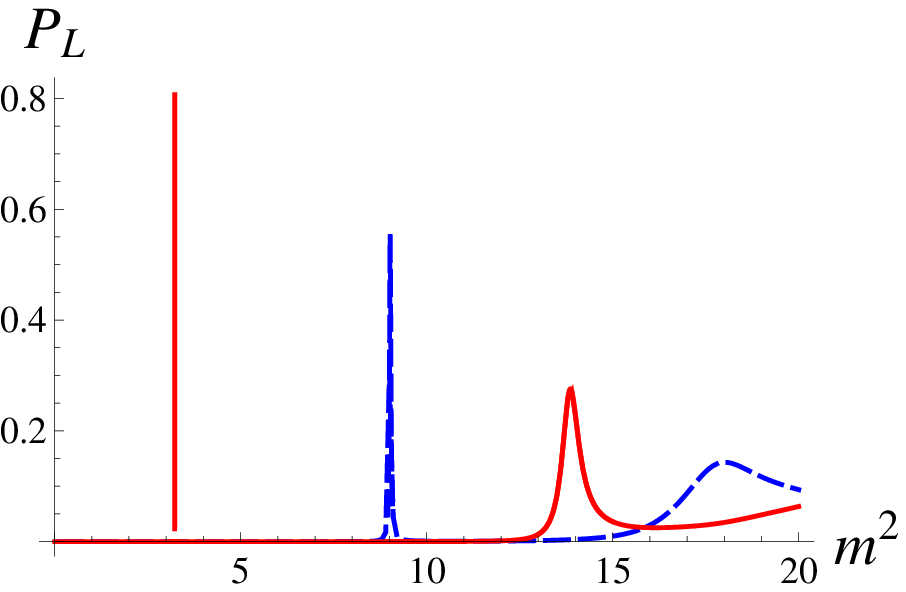}}
    \subfigure[$\eta=-5$]{
    \includegraphics[width=0.22\textwidth]{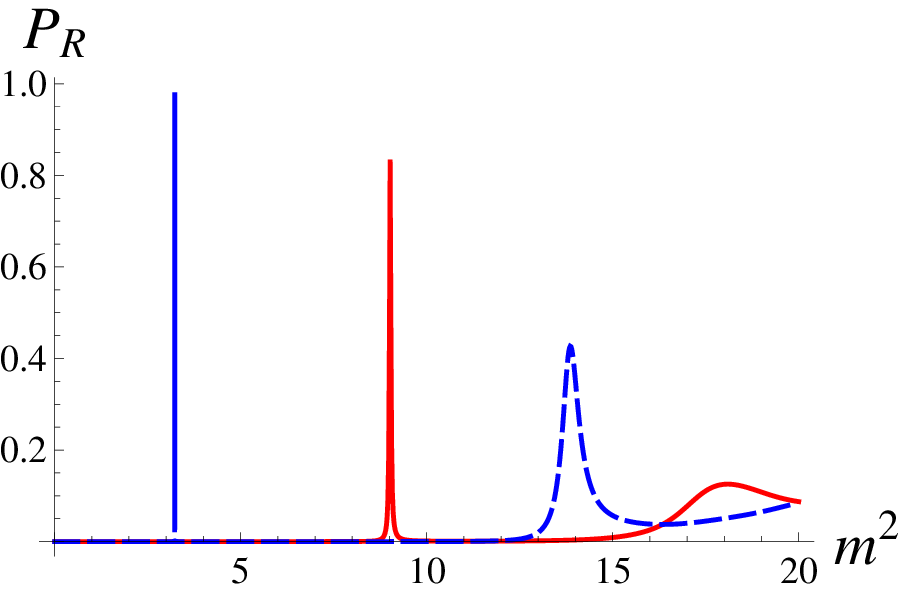}}
    \subfigure[$\eta=-7$]{
    \includegraphics[width=0.22\textwidth]{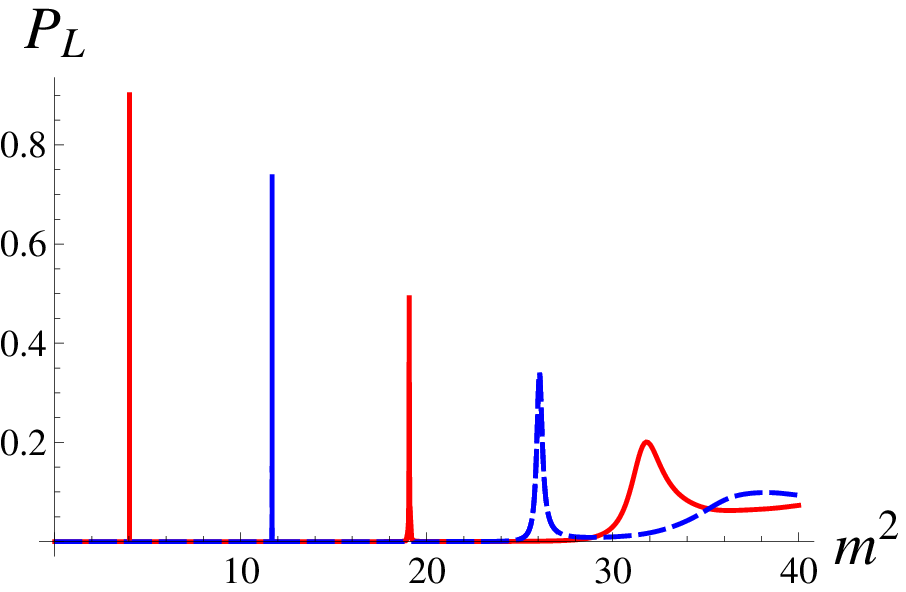}}
    \subfigure[$\eta=-7$]{
    \includegraphics[width=0.22\textwidth]{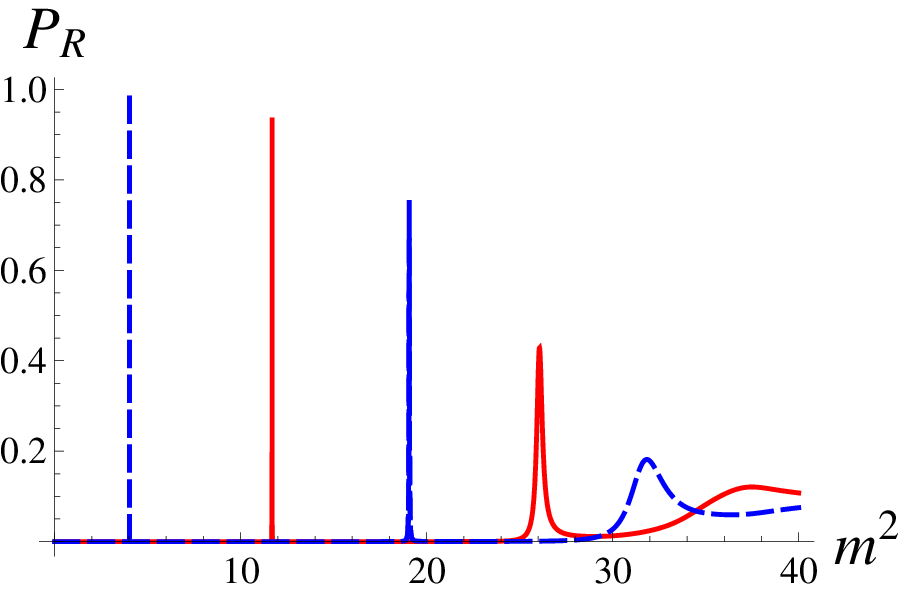}}
    \vskip -4mm \caption{Plots of the probabilities $P_{L,R}$ for the Yukawa coupling mechanism, which include the even parity (blue dashed line) and odd parity (red line) massive KK modes of the left-chiral (up channel) and right-chiral (down channel) fermions.
    The parameters are set to $a=s=1$,  $q=2$, and $\Lambda_{4}=0$.}
    \label{q2peaksl2}
    \end{figure}

We find that, if we fix $a=s=1,~\Lambda_{4}=0,~q=1$ or $q=2$, then there are no resonances for the case of $\eta=-1$. So we only consider the cases of $q=1,~2$ and $\eta=-3,~-5,~-7$. Plots of the relative probability  (\ref{zbmin}) are shown in Figs.~\ref{q1peaksl2} and \ref{q2peaksl2}, which show that the number of the resonant KK fermions increases with $q$ and $|\eta|$. We can obtain the corresponding lifetimes $\tau$ of the fermion resonances by the width ($\Gamma$) at half maximum of the peak with $\tau=\frac{1}{\Gamma}$ \cite{Gregory2000,Almeida2009}. Here, we only list the lifetimes and mass spectrum of the resonances for the case of $q=2$ in Table \ref{TableSpectraYukawa1}. The corresponding resonances for the case of $q=1$ and $\eta=-7$ is shown in Fig.~\ref{ryesonances}.

    \begin{table*}[!htb]
    \begin{center}
    \begin{tabular}{||c|c|c|c|c|c|c|c||}
     \hline
     $\eta$ & $\text{chirality}$ & $\text{parity}$   & $m^{2}_{n}$  & $m_{n}$  & $\Gamma$                    & $\tau$  &  $P$   \\

        \hline \hline

                                         &     & odd   & 2.2336     & 1.4945    & $1.244\times 10^{-2}$    & $80.36 $        & 0.600      \\ \cline{3-8}
        &                  $\mathcal{L}$       & even  & 5.5950     & 2.3654    & $0.156$                  & $6.409 $        & 0.210      \\ \cline{2-8}\cline{2-8}
        \raisebox{2.3ex}[0pt]{-3}        &     & even  & 2.2336     & 1.4945    & $1.261\times 10^{-2}$    & $79.29 $        & 0.927    \\ \cline{3-8}
                          &$\mathcal{R}$       & odd   & 5.5825     & 2.3622    & $0.176$                  & 5.675           & 0.396      \\
        \hline\hline
                              &                & ood   & 3.2302     & 1.7973    & $2.058\times 10^{-4}$    & $4857$          & 0.811    \\  \cline{3-8}
                  & $\mathcal{L}$              & even  & 9.0201     & 3.0033    & $7.491\times 10^{-3}$    & 133.5           & 0.552    \\  \cline{3-8}
                              &                & odd   & 13.865     & 3.7236    & $0.0762$                 & 13.13           & 0.276    \\  \cline{3-8}\cline{2-8}
      \raisebox{2.3ex}[0pt]{-5}            &   & even  & 3.2302     & 1.7973    & $1.975\times 10^{-4}$    & $5063$          & 0.981    \\  \cline{3-8}
                              & $\mathcal{R}$  & odd  & 9.0200     & 3.0033    & $7.491\times 10^{-3}$    & 133.5           & 0.835    \\  \cline{3-8}
                                           &   & even   & 13.8700    & 3.7242    & $0.0751$                 & 13.31           & 0.427    \\  \cline{3-8}
      \hline\hline

                                           &   & odd   & 4.0289     & 2.0072   & $1.171\times10^{-6}$      & $8.541\times 10^{5}$      & 0.906   \\  \cline{3-8}
                              & $\mathcal{L}$  & even  & 11.6927    & 3.4195   & $1.535\times10^{-4}$      & $6513$                    & 0.740   \\  \cline{3-8}
                              &                & odd   & 19.0650    & 4.3664   & $3.264\times10^{-3}$      & 306.4                      & 0.497   \\  \cline{3-8}
                              &                & even  & 26.0660    & 5.1059   & $0.0352$                  & 28.37                     & 0.340   \\  \cline{2-8}
      \raisebox{2.3ex}[0pt]{-7}            &   & even  & 4.0289     & 2.0072   & $1.196\times10^{-6}$      & $8.363\times 10^{5}$      & 0.994   \\  \cline{3-8}
                              & $\mathcal{R}$  & odd   & 11.6927    & 3.4195   & $1.550\times10^{-4}$      & $6452$         & 0.938   \\  \cline{3-8}
                              &                & even  & 19.0650    & 4.3664   & $3.206\times10^{-3}$      & 311.9           & 0.750   \\  \cline{3-8}
                              &                & odd  & 26.0670    & 5.1056    & $0.0351$                  & 28.53                      & 0.429   \\  \cline{3-8}
       \hline

    \end{tabular}\\
    \caption{The mass spectrum $m^2_n$ and $m_n$, width ($\Gamma$), lifetime ($\tau$), and relative probability ($P$) of the left- and right-chiral KK fermion resonances for the Yukawa coupling. The parameters are set to $q=2,~a=s=1,~\Lambda_{4}=0$. The symbols $\mathcal {L}$ and $\mathcal {R}$ are short for left-chiral and right-chiral, respectively.}
    \label{TableSpectraYukawa1}
    \end{center}
    \end{table*}

\subsubsection{\textbf{LXCW coupling mechanism}}

In this subsection, we consider the LXCW coupling mechanism reviewed in Sec.~\ref{review}. Since the coupling function $F_1$ should be an even function of the extra dimension, we choose $F_1$ as $F_1=\text{arcsinh}^{2p}(b\phi)$ with a structure parameter $p$ (positive integer). Then the effective potentials are given by

  \begin{eqnarray}
    V_{L,R}^{\text{new}}(z)
     \!\!&=&\!\! \pm\frac{2p\eta h^2 \text{arcsinh}^{2p-2}(hz)}
                 {(1+h^2 z^2 )^{3/2}}   \nonumber\\
    && \times\bigg((2 p-1) \sqrt{1+h^2z^2}-hz~\text{arcsinh}(hz)\nonumber\\
    && \pm2p\eta\sqrt{1+h^2z^2}~\text{arcsinh}^{2 p}(hz)\bigg).
    \label{potentialnewz}
  \end{eqnarray}

\begin{figure}[!htb]
    \subfigure[$p=1$]{
    \includegraphics[width=0.22\textwidth]{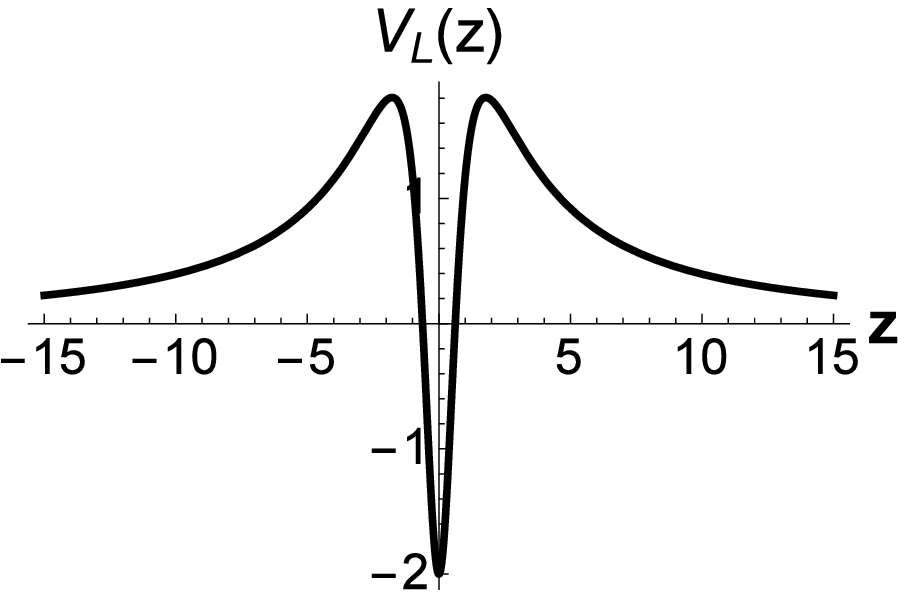}}
    \subfigure[$p=1$]{
    \includegraphics[width=0.22\textwidth]{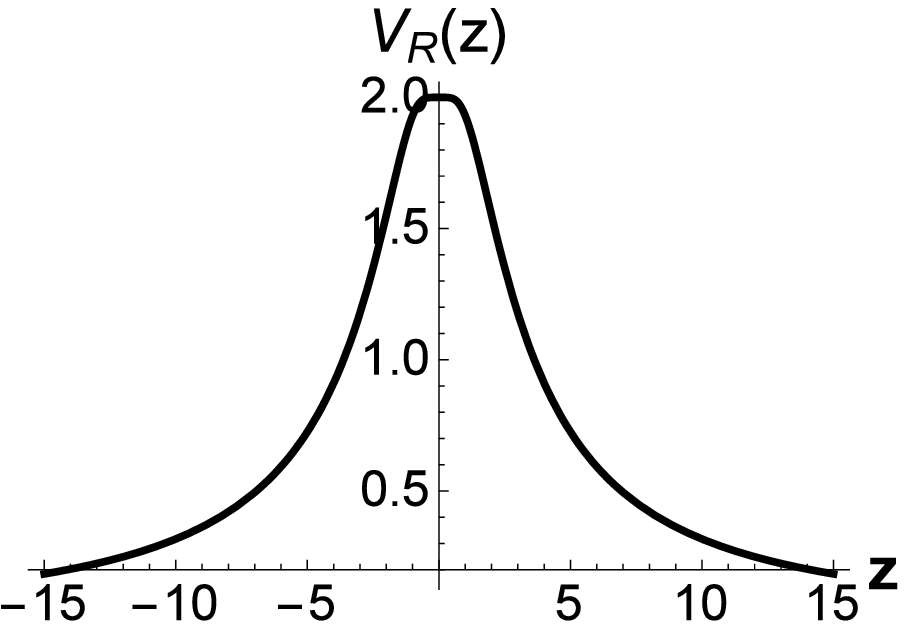}}
    \subfigure[$p=2$]{
    \includegraphics[width=0.22\textwidth]{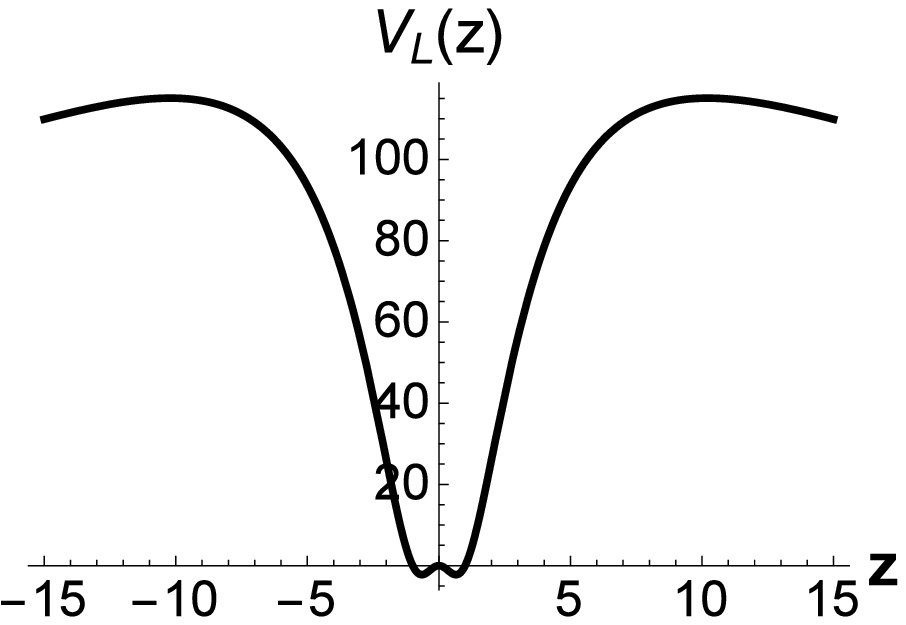}}
    \subfigure[$p=2$]{
    \includegraphics[width=0.22\textwidth]{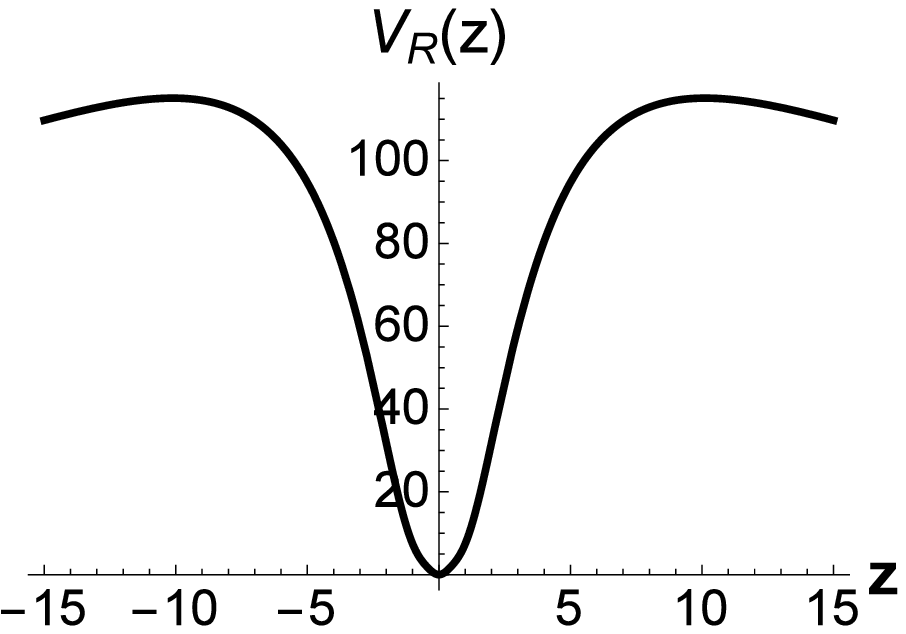}}
    \subfigure[$p=3$]{
    \includegraphics[width=0.22\textwidth]{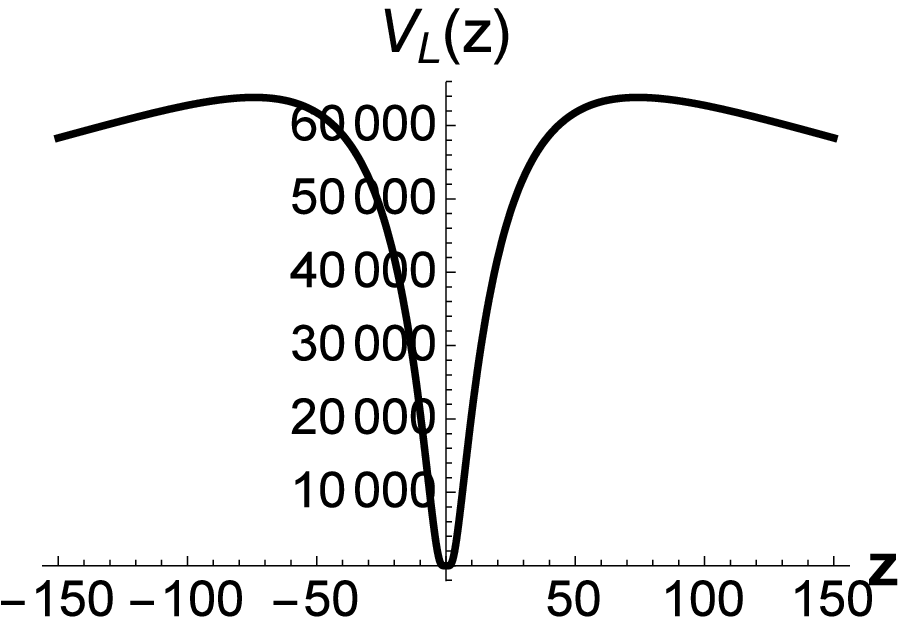}}
    \subfigure[$p=3$]{
    \includegraphics[width=0.22\textwidth]{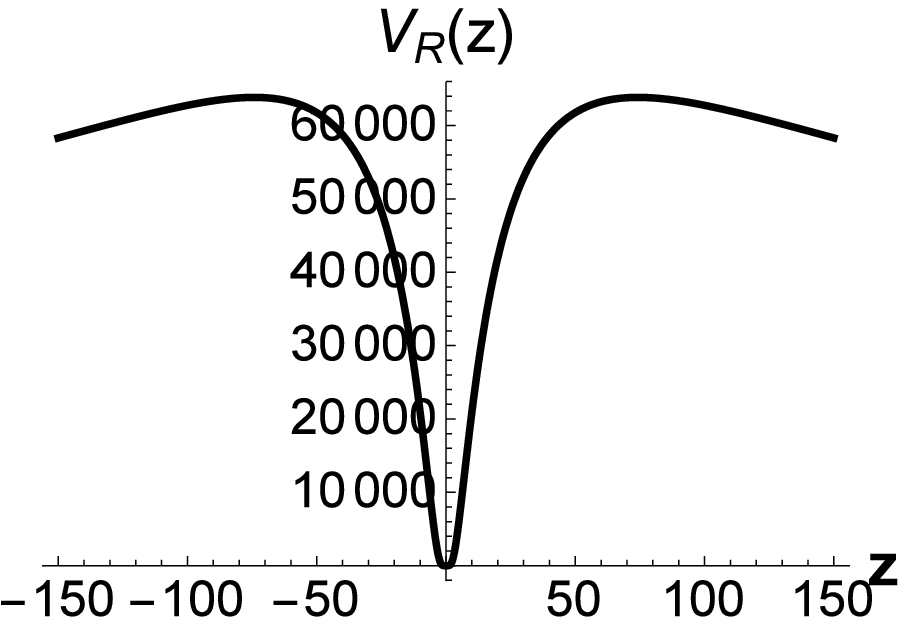}}
    \vskip -4mm \caption{Plots of the effective potentials (\ref{potentialnewz}) with the LXCW coupling  for different values of $p$ and $\eta=-1$.}
    \label{potentialchange}
    \end{figure}

\begin{figure}[!htb]
    \includegraphics[width=0.22\textwidth]{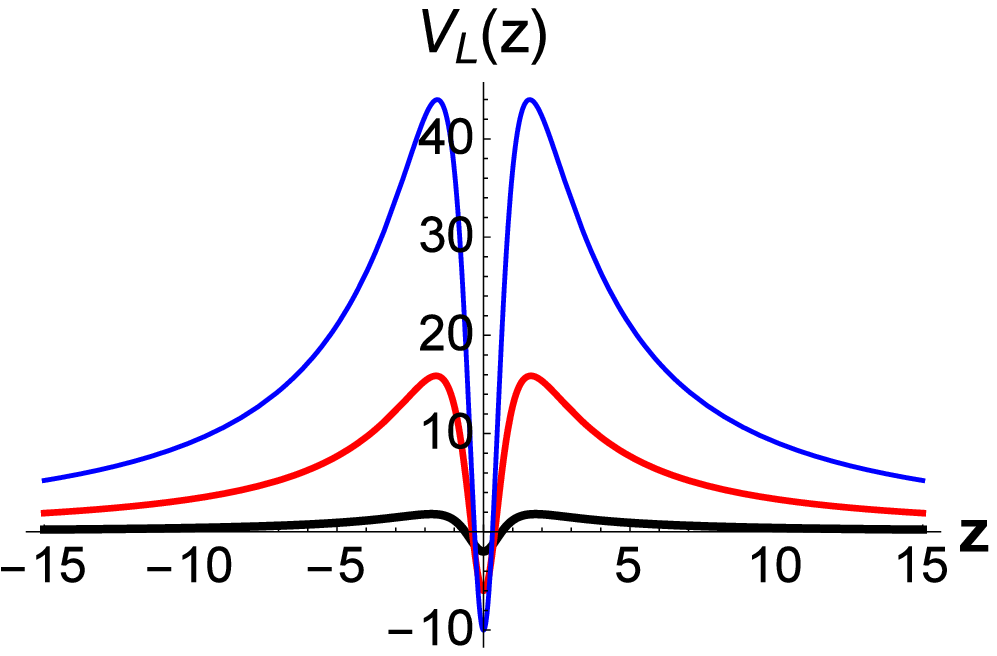}~~~~
    \includegraphics[width=0.22\textwidth]{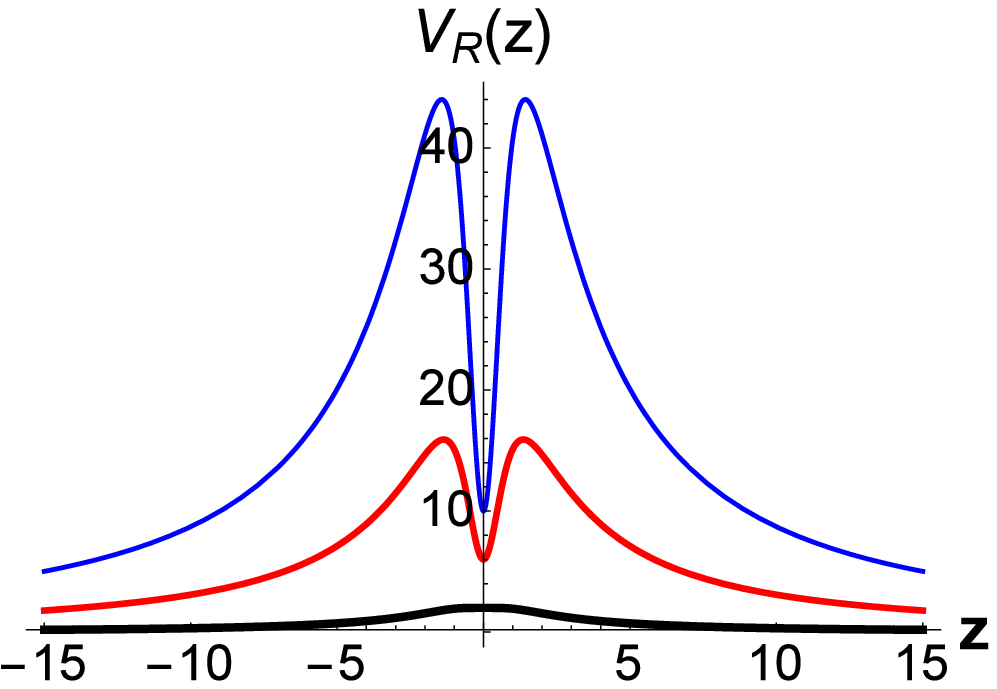}
    \vskip -4mm \caption{Plots of the effective potentials (\ref{potentialnewz}) with the LXCW coupling  for $p=1$ and different values of $\eta$ ($-1$ for thick black line, $-3$ for red line, and $-5$ for thin blue line).}
    \label{potentialchange1}
    \end{figure}

By defining $\bar{z}=hz$, $\overline{V}_{L,R}^{\text{new}}=h^{-2}V_{L,R}^{\text{new}}$, and $\bar{\mu}={\mu}/h$, the Schr\"{o}dinger-like equations (\ref{schrodingerlikeequationl}) and  (\ref{schrodingerlikeequationr}) become \begin{eqnarray}
    \left[-\partial_{\bar{z}}^2+\overline{V}_{L,R}(\bar{z})\right]f_{Ln,Rn}(\bar{z}) =\bar{\mu}^2_nf_{Ln,Rn}(\bar{z}). \label{schrodingerlikeequationsNew}
    \end{eqnarray}
Therefore, the role of the parameter $h$ is just a rescaling of the mass spectrum. Thus, we take $h=1$ without loss of generality. The asymptotic behaviors of the above effective potentials are described as follows
    \begin{eqnarray}
        V_{L,R}^{\text{new}}(0) &=&\left\{\begin{array}{cl}
                                  \pm2\eta, & ~~~p=1 \\
                                  0, & ~~~p\ge2
                                \end{array} \right.
                                \label{Vasymptotic1} \\
        V_{L,R}^{\text{new}}(\infty)&\to& 0.\label{Vasymptotic2}
    \end{eqnarray}
Now it is clear that for the LXCW coupling mechanism with $F_1=\arctan^{2p}(b\phi)$, the parameters $a$, $b$, $s$, and $\Lambda_{4}$ in the single-scalar-field thick brane model do not affect the structures of the effective potentials (\ref{potentialnewz}), and only $\eta$ and $p$ have affect.
Plots of the effective potentials with different values of coupling parameters $p$ and $\eta$ are shown in Figs. \ref{potentialchange} and \ref{potentialchange1}, respectively. It can be seen from Fig. \ref{potentialchange} that there will appear a quasi-well for $V_R$ when $p\ge2$ and the potential barriers for $V_{L,R}$ will increase quickly with the parameter $p$. The large coupling ($-\eta$) also results in a quasi-well for $V_R$ and large potential barriers for $V_{L,R}$ (see Fig. \ref{potentialchange1}).

The solution of the zero modes are
    \begin{eqnarray}
    f_{L0,R0}^{\text{new}}(z)
     &=&  N_{L,R} \exp(\pm\eta F_1) \nonumber \\
     &=&  N_{L,R}  \exp\left(\pm\eta~\text{arcsinh}^{2 p}(hz)\right),
    \label{firstzeromode}
    \end{eqnarray}
where $N_{L,R}$ are normalization constants.
Since $\text{arcsinh}^{2p}(z)\rightarrow\infty$ at $z\to\pm\infty$, in order to localize the massless left-chiral fermion on the brane, the parameter $\eta$ should be negative, for which the massless right-chiral fermion can not be localized. The normalization condition (\ref{orthonormality}) for the massless left-chiral mode is
    \begin{eqnarray}
    \int_0^\infty{|f_{L0}(z)|^2d{z}}  <\infty \label{zeromodelocalizedcondition}
    \end{eqnarray}
is satisfied because
    \begin{eqnarray}
    |f_{L0}(z\rightarrow\infty)|^2
    %&\propto& \exp\left[2\eta~\text{arcsinh}^{2 p}(hz)\right] \nonumber\\
    &\rightarrow& N_{L} \exp\left[2\eta~\ln^{2 p}(hz)\right]  \nonumber\\
    &=& N_{L}\left(\frac{1}{hz}\right)^{-2\eta \ln^{2 p-1}(hz)} \nonumber \\
    &\rightarrow & N_{L}\left(\frac{1}{hz}\right)^{+\infty}.     \label{firstzeromode1}
    \end{eqnarray}
Which means that we can get the four-dimensional massless left-chiral fermion on the thick brane. From the asymptotic behaviors (\ref{Vasymptotic1}) and (\ref{Vasymptotic2}), we require $V_{L}^{\text{new}}(0)<0$ in order to localize the massless left-chiral fermion. Therefore, the parameter $\eta$ should be negative, which is consistent with the above result (\ref{zeromodelocalizedcondition}).

    \begin{figure}[!htb]
    \subfigure[$\eta=-1$]{
    \includegraphics[width=0.22\textwidth]{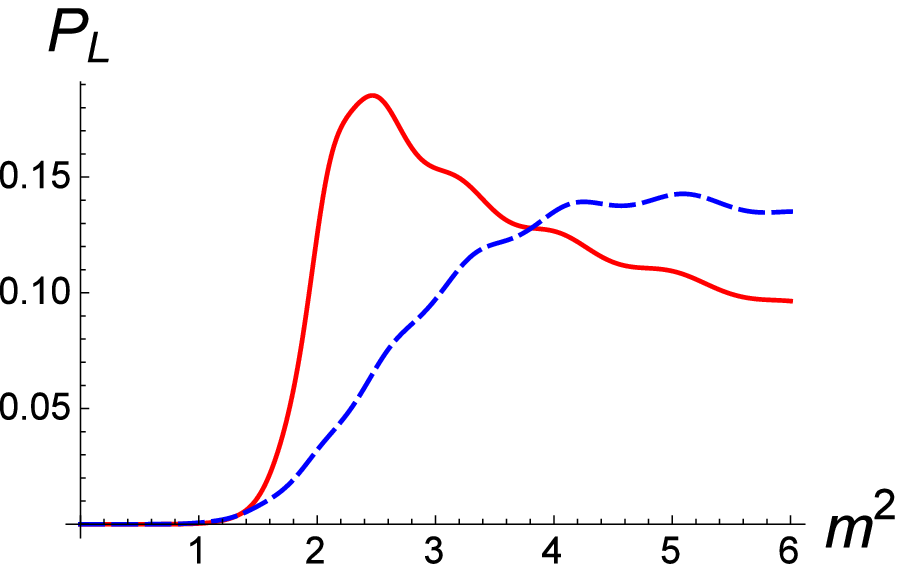}}
    \subfigure[$\eta=-1$]{
    \includegraphics[width=0.22\textwidth]{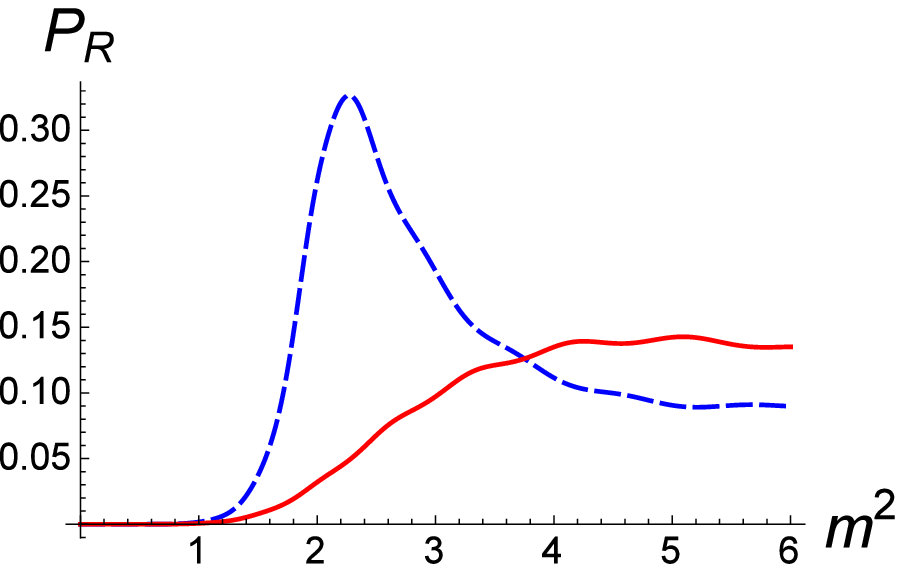}}
    \subfigure[$\eta=-3$]{
    \includegraphics[width=0.22\textwidth]{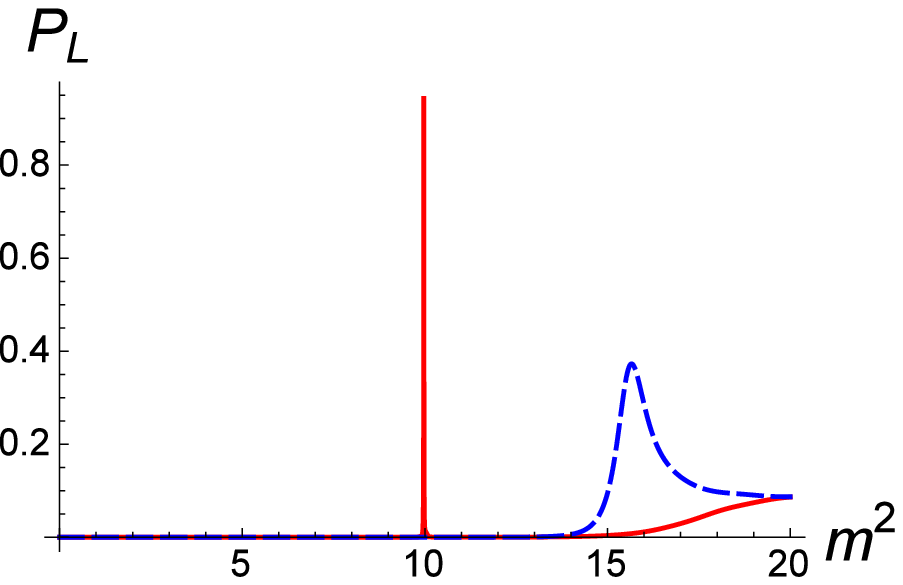}}
    \subfigure[$\eta=-3$]{
    \includegraphics[width=0.22\textwidth]{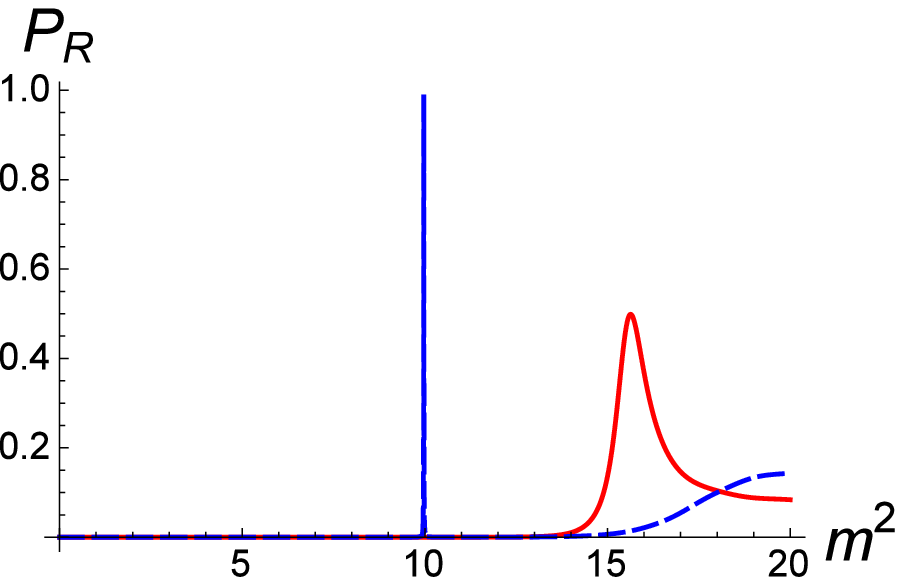}}
    \subfigure[$\eta=-5$]{
    \includegraphics[width=0.22\textwidth]{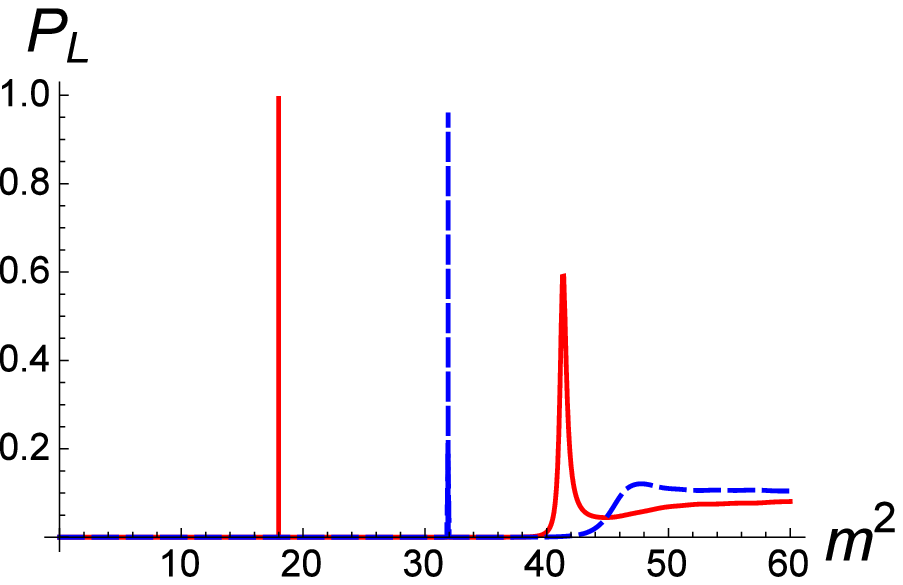}}
    \subfigure[$\eta=-5$]{
    \includegraphics[width=0.22\textwidth]{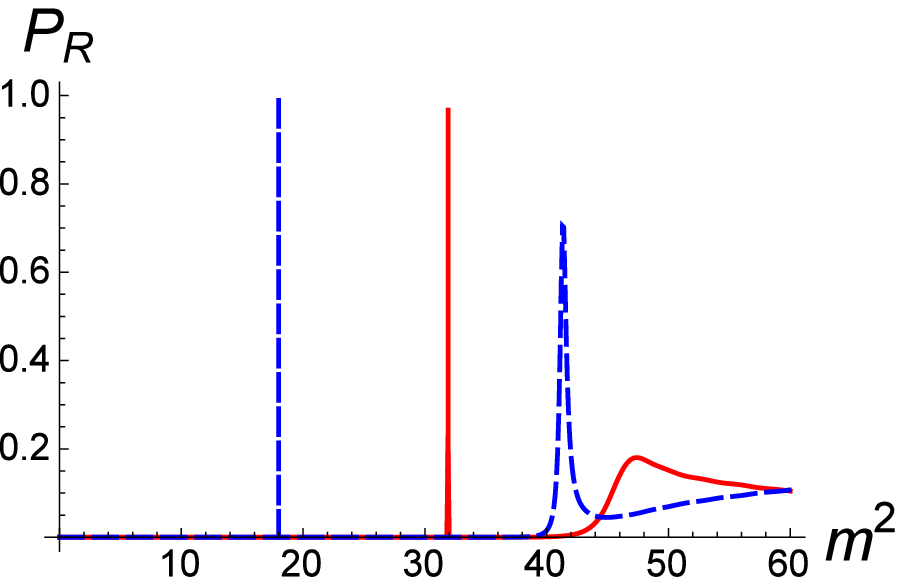}}
    \vskip -4mm \caption{Plots of the probabilities $P_{L,R}$ for the LXCW coupling mechanism, which include even parity (blue dashed line) and odd parity (red line) massive KK modes of the left-chiral (up channel) and right-chiral (down channel) fermions. The parameters are set to $a=s=1$, $p=1$, and $\Lambda_{4}=0$.}
    \label{peaksl1}
    \end{figure}

\begin{figure}[!htb]
    \subfigure[$f_{L1}(z)$]{
    \includegraphics[width=0.22\textwidth]{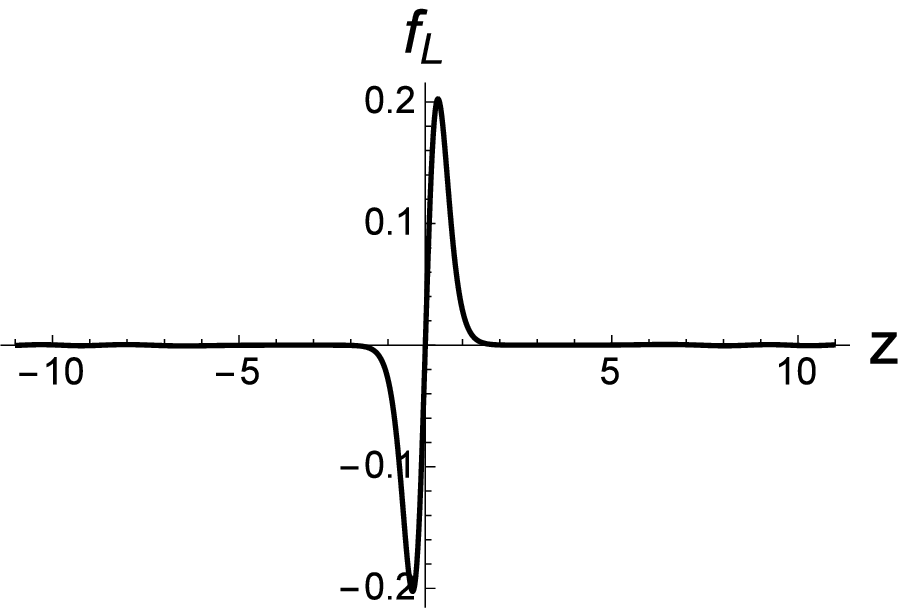}}
    \subfigure[$f_{R1}(z)$]{
    \includegraphics[width=0.22\textwidth]{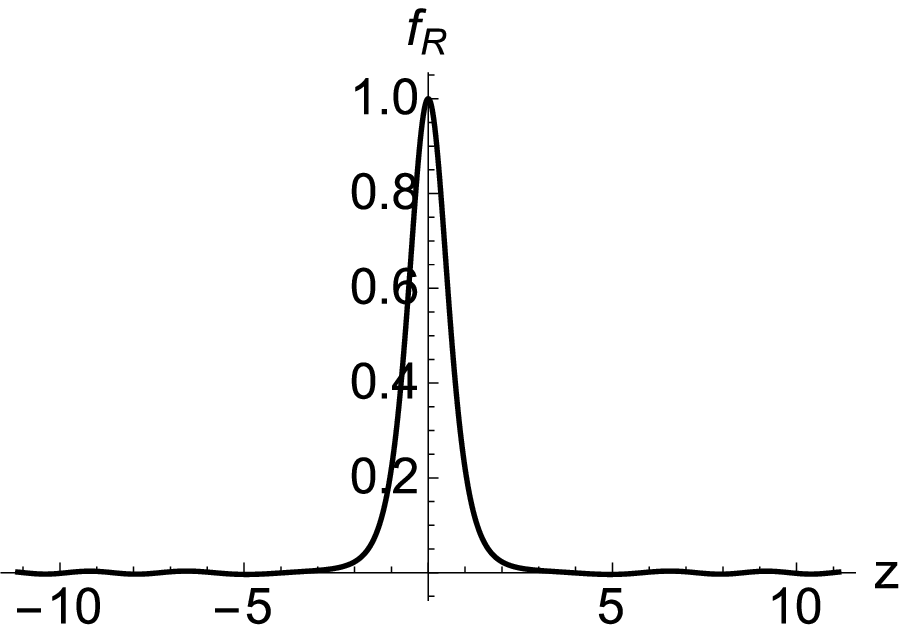}}
    \subfigure[$f_{L2}(z)$]{
    \includegraphics[width=0.22\textwidth]{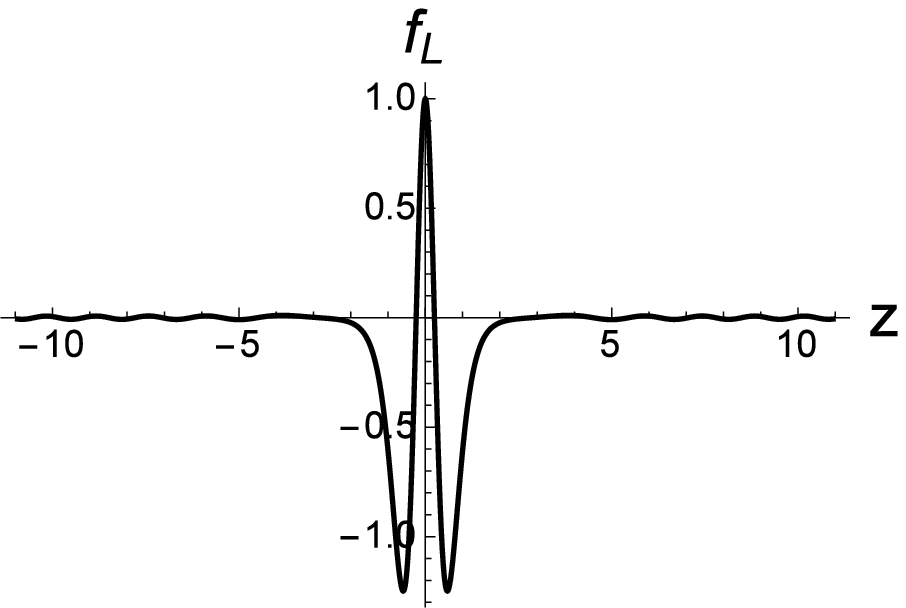}}
    \subfigure[$f_{R2}(z)$]{
    \includegraphics[width=0.22\textwidth]{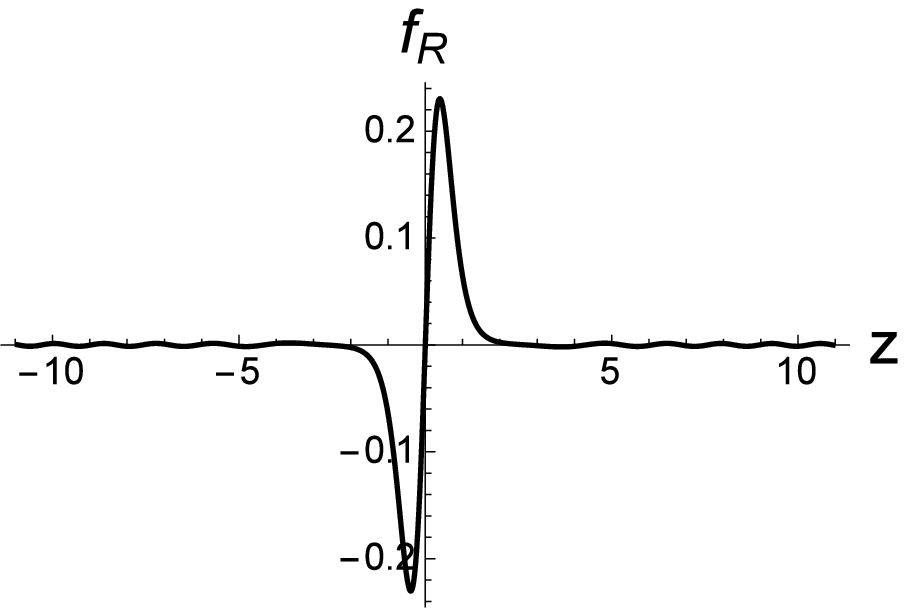}}
    \subfigure[$f_{L3}(z)$]{
    \includegraphics[width=0.22\textwidth]{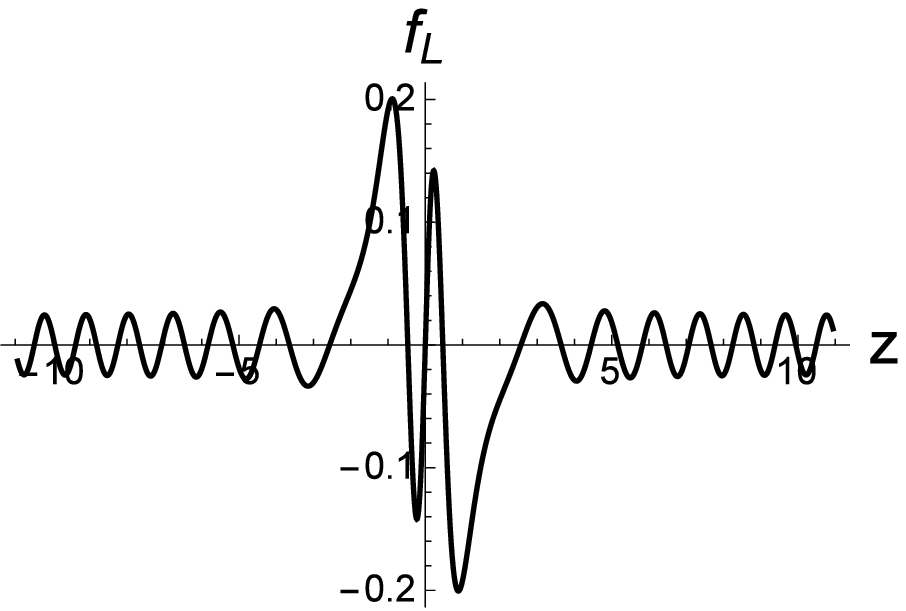}}
    \subfigure[$f_{R3}(z)$]{
    \includegraphics[width=0.22\textwidth]{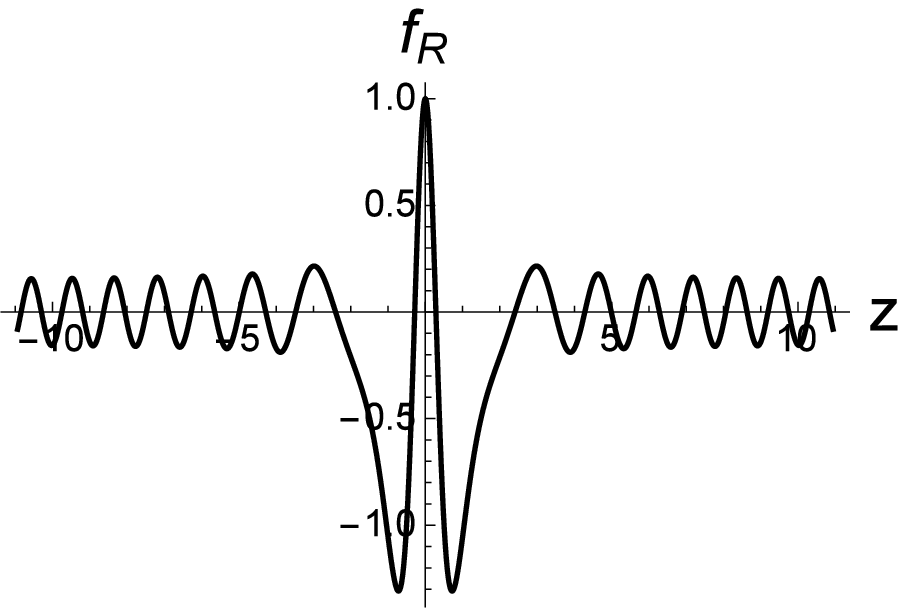}}
    \vskip -4mm \caption{Plots of the resonances of the left- and right-chiral KK fermions for the LXCW coupling mechanism with $\eta=-5$ and $p=1$.}
    \label{resonances}
    \end{figure}

Because the depth of the potentials increases quickly with the parameter $p$ for fixed coupling constant $\eta$, the number of massive KK modes are also increasing quickly, in the following we mainly consider the effect of $\eta$ on the massive KK modes with $p=1$. Plots of the relative probability with different values of $\eta$ and the corresponding wave functions with $\eta=-5$ are shown in Figs.~\ref{peaksl1} and \ref{resonances}. Furthermore, we obtain the corresponding mass spectrum of KK modes, and the lifetimes $\tau$ of fermion resonances can be calculated by the width ($\Gamma$) at half maximum of the peak with $\tau=\frac{1}{\Gamma}$ \cite{Gregory2000,Almeida2009}, which are listed in TABLE \ref{TableSpectranew}. Because the precision of the program wolfram mathematica is not enough we do not give the lifetime of KK mode with $m^2=27.996951884884$ and $\eta=-5$.

    \begin{table*}[!htb]
    \begin{center}%\renewcommand\arraystretch{1.4}
    \begin{tabular}{||c|c|c|c|c|c|c|c||}
     \hline
     $\eta$ & chirality & parity   & $m^{2}_{n}$  & $m_{n}$  & $\Gamma$                    & $\tau$  &  $P$   \\

        \hline \hline

                                         &     & odd   & 9.9697    & 3.1575   & $1.172\times10^{-3}$        & 853.37               & 0.948     \\ \cline{3-8}
        &                  $\mathcal{L}$       & even  & 15.6585   & 3.9573   & $0.150$                     & 6.684                & 0.373    \\
        \cline{2-8}\cline{2-8}
        \raisebox{2.3ex}[0pt]{-3}        &     & even  & 9.9697    & 3.1575   & $1.172\times10^{-3}$        & 853.37               &0.987      \\ \cline{3-8}
                          &$\mathcal{R}$       & odd   & 15.6327   & 3.9535   & $0.142$                     & 7.029                &0.499     \\
    \hline\hline

                                           &   & odd   & 17.9915   & 4.2416   & $9.666\times10^{-9}$        & $1.035\times 10^8$    & 0.999   \\  \cline{3-8}
                              & $\mathcal{L}$  & even  & 31.9057   & 5.6485   & $2.656\times10^{-4}$        & $3765 $               & 0.961   \\  \cline{3-8}
                              &                & odd   & 41.3497   & 6.4304   & $0.050$        & 20.10                & 0.605  \\  \cline{2-8}
      \raisebox{2.3ex}[0pt]{-5}            &   & even  & 17.9915   & 4.2416   & $9.548\times10^{-9}$        & $1.047\times 10^8$     & 0.999   \\  \cline{3-8}
                              & $\mathcal{R}$  & odd   & 31.9057   & 5.6485   & $4.957\times10^{-4}$        & $2017$     & 0.987   \\  \cline{3-8}
                              &                & even  & 41.3471   & 6.4302   & $0.042$        & 23.83                 & 0.721   \\  \cline{3-8}
       \hline

    \end{tabular}\\
    \caption{The mass spectrum $m^2_n$ and $m_n$, width ($\Gamma$), lifetime ($\tau$), and relative probability ($P$) of the left- and right-chiral KK fermion resonances for the LXCW coupling. The parameters are set to $p=1,~a=s=1$, and $\Lambda_{4}=0$. }
    \label{TableSpectranew}
    \end{center}
    \end{table*}

Before closing this subsection, we give a brief summary. We have considered localization and resonances of left- and right-chiral fermions on a single-scalar-field thick brane with two kinds of scalar-fermion couplings. The mass spectra and lifetimes of the fermion resonances for both the left- and right-chiral fermions are the same. The number of the fermion resonances increases with the scalar-fermion coupling parameter $|\eta|$ and the brane structure parameter.

\subsection{Multi-scalar-field flat thick brane}

\begin{figure}[!htb]
    \subfigure[$a=0.02$]{
    \includegraphics[width=0.22\textwidth]{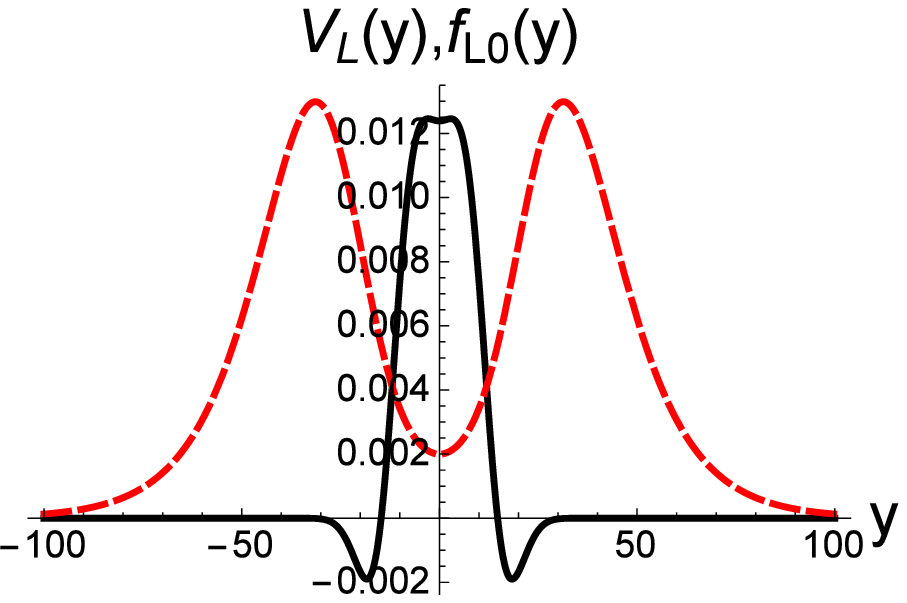}}
    \subfigure[$a=0.02$]{
    \includegraphics[width=0.22\textwidth]{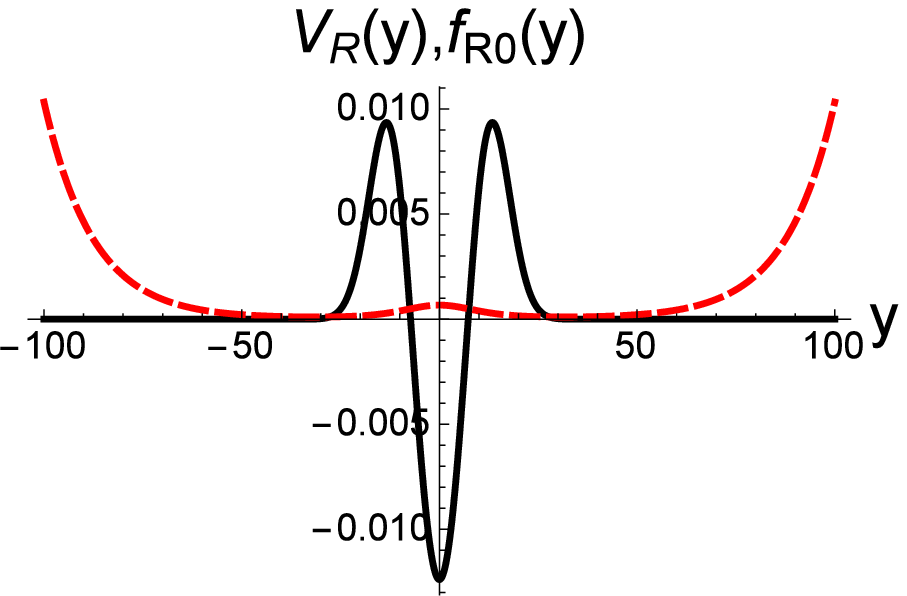}}
    \subfigure[$a=0.35$]{
    \includegraphics[width=0.22\textwidth]{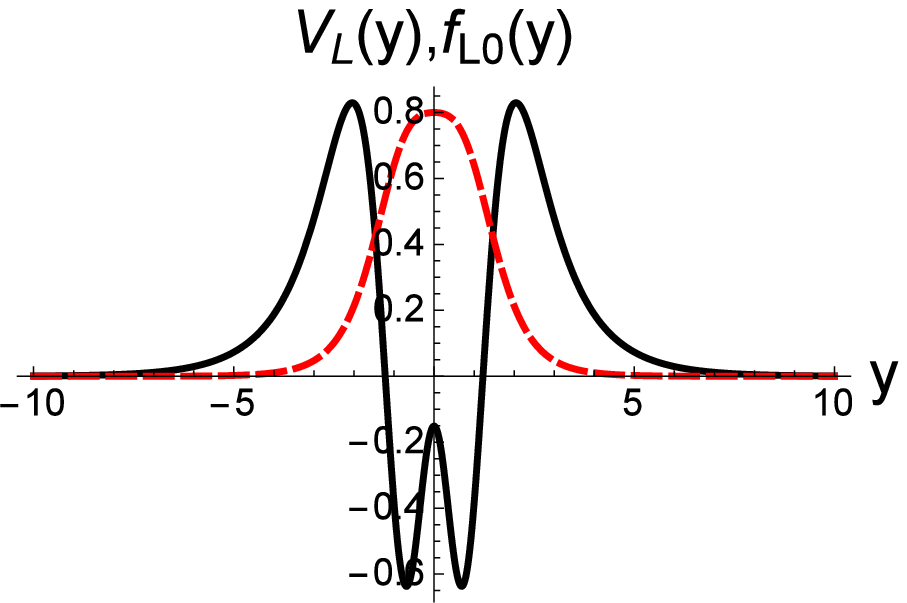}}
    \subfigure[$a=0.35$]{
    \includegraphics[width=0.22\textwidth]{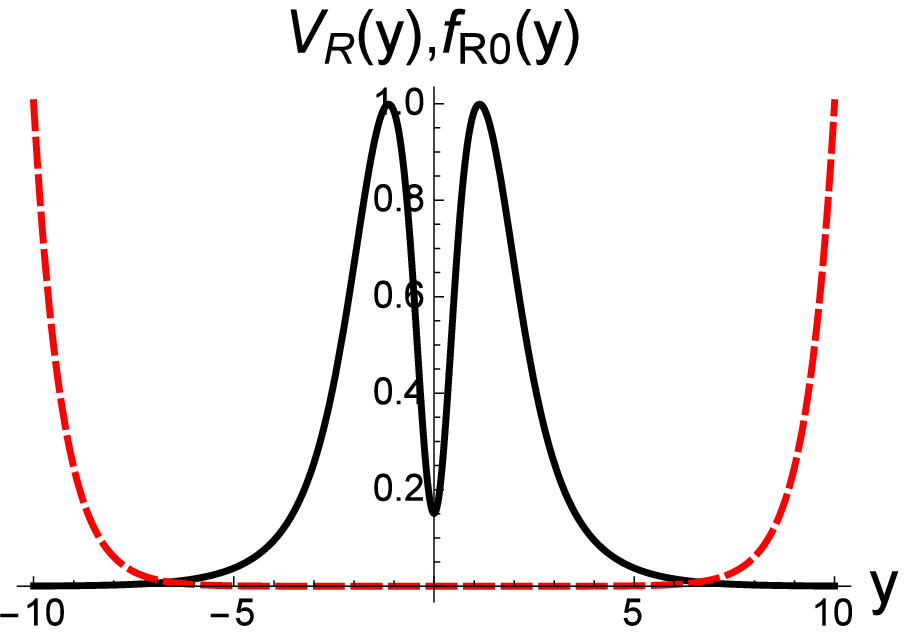}}
    \subfigure[$a=0.45$]{
    \includegraphics[width=0.22\textwidth]{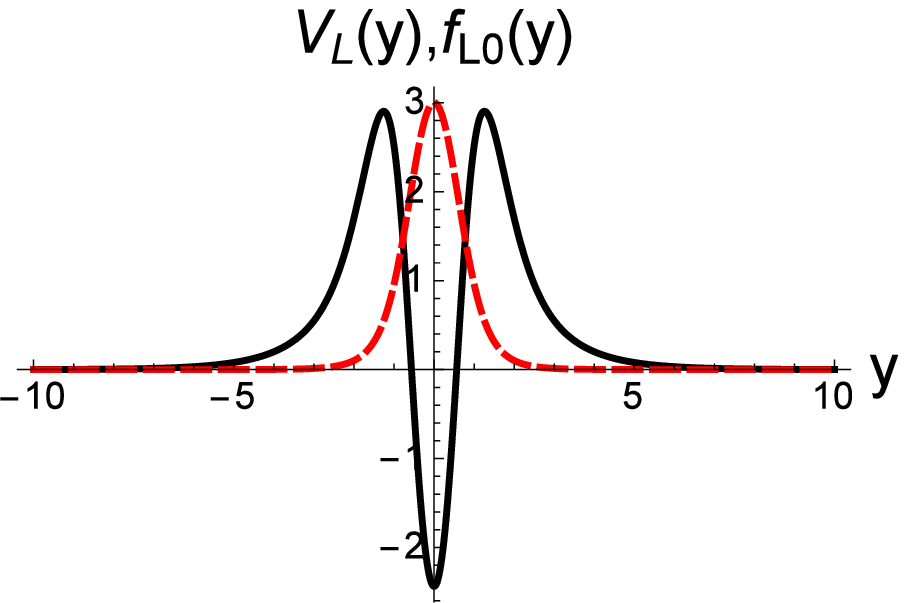}}
    \subfigure[$a=0.45$]{
    \includegraphics[width=0.22\textwidth]{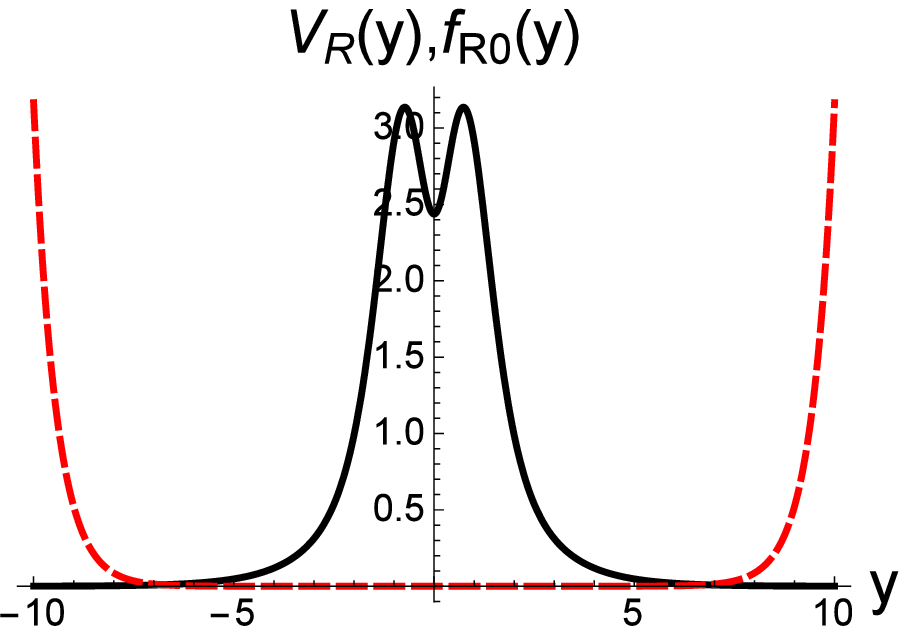}}
    \vskip -4mm \caption{Plots of the effective potentials (black line) and zero modes (dashed red line) for the multi-scalar-field model with different values of the parameter $a$. The parameters are set to $p=1$ and $\eta=1$.}
    \label{figchangea}
    \end{figure}

\begin{figure}[!htb]
    \subfigure[$p=1$]{
    \includegraphics[width=0.22\textwidth]{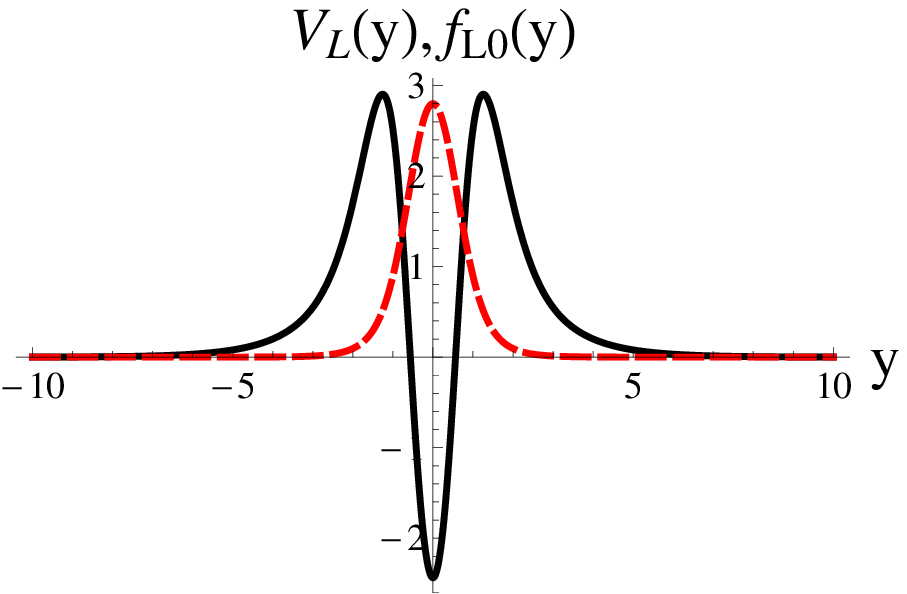}}
    \subfigure[$p=1$]{
    \includegraphics[width=0.22\textwidth]{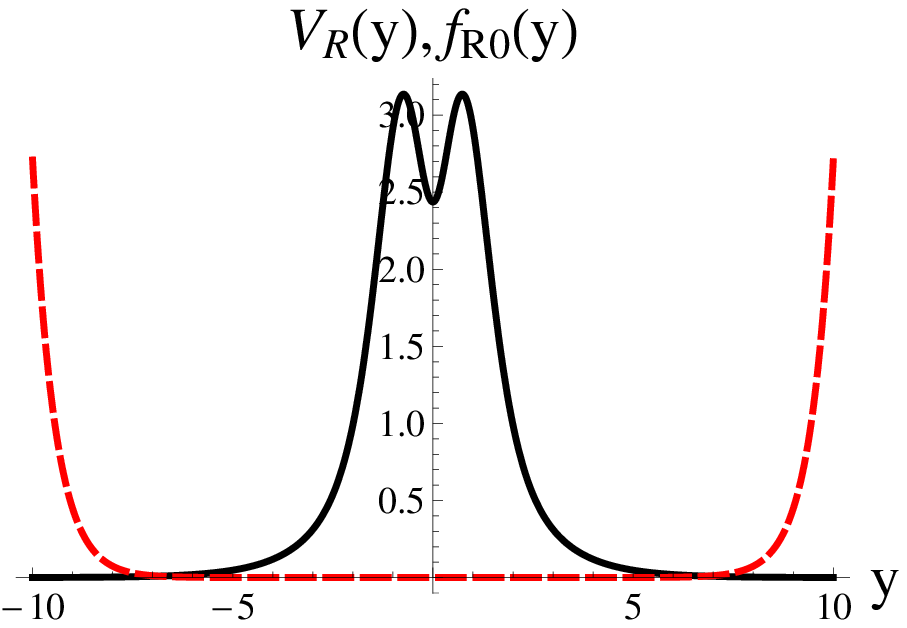}}
    \subfigure[$p=2$]{
    \includegraphics[width=0.22\textwidth]{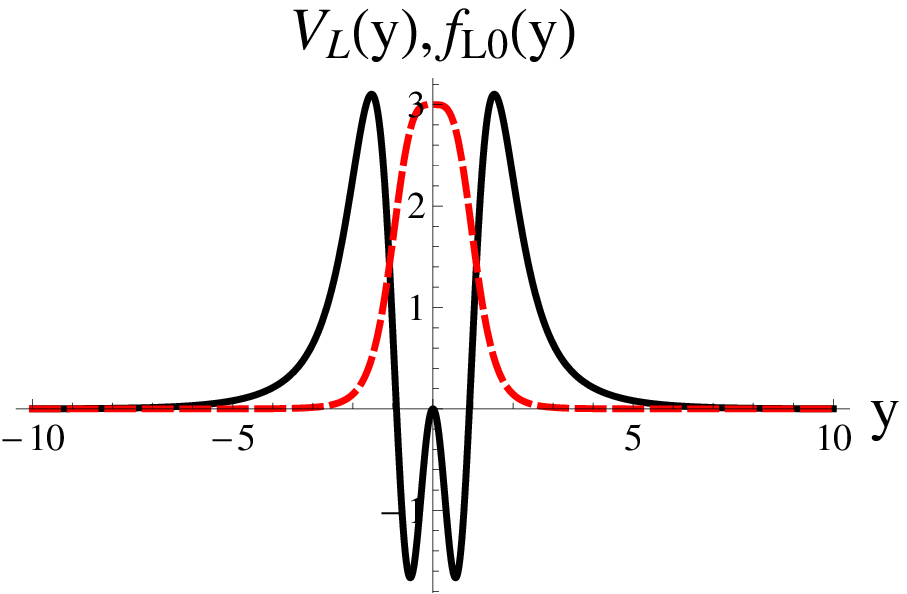}}
    \subfigure[$p=2$]{
    \includegraphics[width=0.22\textwidth]{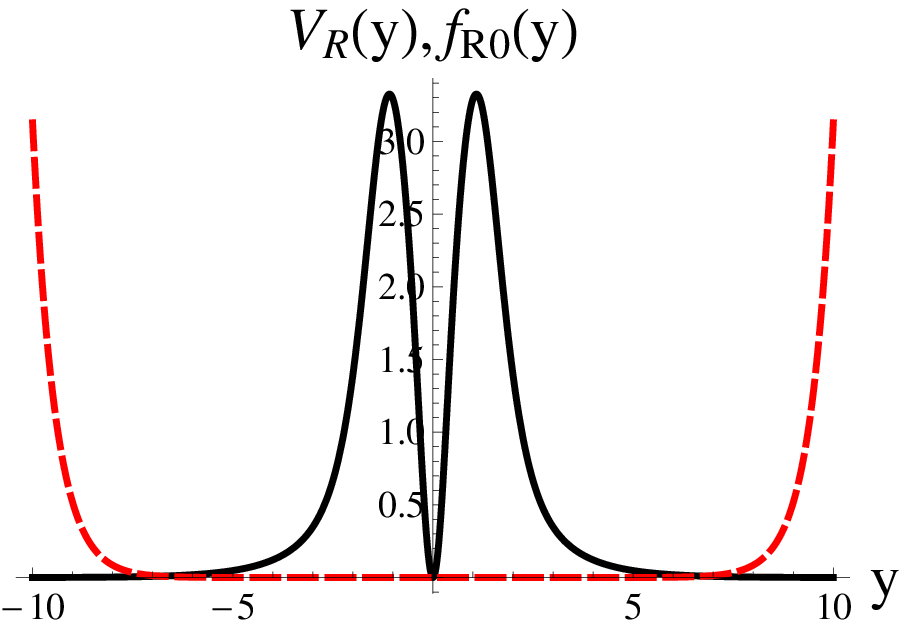}}
    \subfigure[$p=3$]{
    \includegraphics[width=0.22\textwidth]{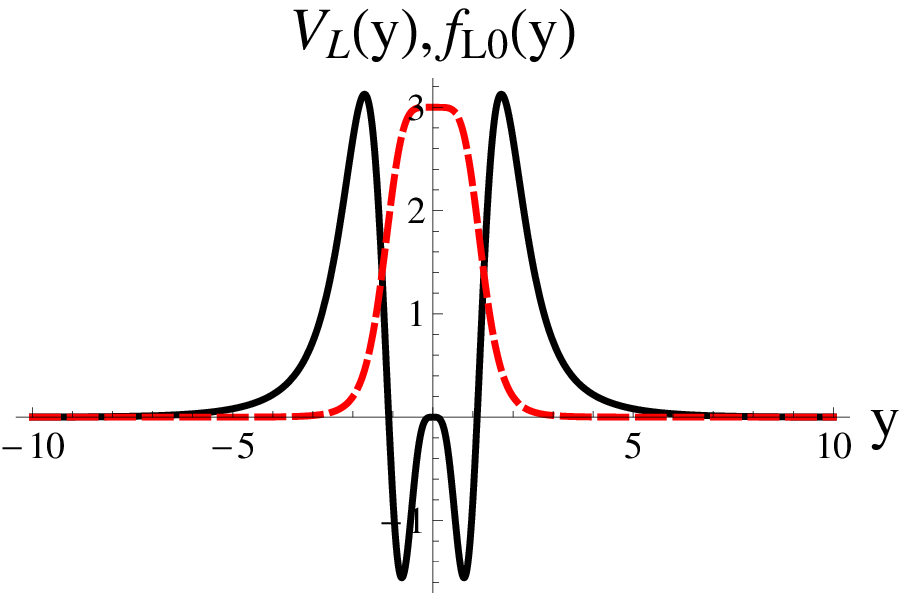}}
    \subfigure[$p=3$]{
    \includegraphics[width=0.22\textwidth]{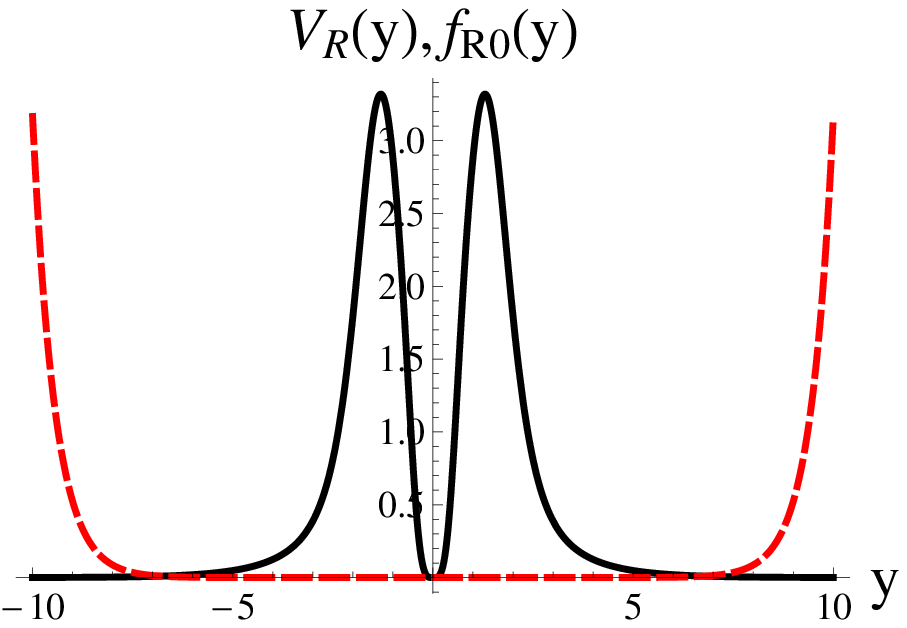}}
    \vskip -4mm \caption{Plots of the effective potentials (black line) and zero modes (dashed red line) for the second model with different values of $p$. The parameters are set to $\eta=1$ and $a=0.45$.}
    \label{potentialchangep}
    \end{figure}

In this subsection, we mainly investigate localization and resonances of fermion based on the new (LXCW) coupling mechanism on a multi-scalar-field flat thick brane embedded in a five-dimensional  spacetime. The action for the brane system is given by \cite{Bazeia2002,Liu2009}
    \begin{eqnarray}
    S&=&\int d^4x~dy\sqrt{-g}\bigg(\frac{M_5^3}{4}R-\frac{1}{2}\partial_M\phi\partial^M\phi\nonumber\\
       &-&\frac{1}{2}\partial_M\chi\partial^M\chi-\frac{1}{2}\partial_M\rho\partial^M\rho-V(\phi,\chi,\rho)\bigg).
    \end{eqnarray}
We also set $M_5=1$. The scalar potential $V$ is taken as the following form \cite{Bazeia2002}
    \begin{eqnarray}
    V&=&\frac{1}{2}\bigg[ 4a^2\phi^2(\rho^2+\chi^2)
                      +\left[1-\phi^2+a(\rho^2+\chi^2)\right]^2
               \bigg]\nonumber\\
      &-&\frac{4}{3}\phi^2\left[1-\frac{1}{3}\phi^2-a(\chi^2+\rho^2)
                \right]^2.
    \end{eqnarray}
A brane solution was found in Refs. \cite{Bazeia2002}:
    \begin{eqnarray}
    \phi(y)&=&\tanh(ky),\nonumber\\
    \chi(y)&=&\sqrt{\frac{1}{a}-2}\cos(\theta) \text{sech}(ky),\nonumber\\
    \rho(y)&=&\sqrt{\frac{1}{a}-2}\sin(\theta) \text{sech}(ky),\nonumber\\
    A(y)&=&\frac{1}{9a}\left[(1-3a)\tanh^2(ky)-2\ln\cosh(ky)\right],
    \end{eqnarray}
where $k=2a$, the domains of parameters $a$ and $\theta$ are $0<a<1/2$ and $\theta\in[0,2\pi)$. For the kink configuration of the scalar $\phi(y)$ and the lump configurations of $\chi(y)$ and $\rho(y)$, we choose the function $F_1=\phi^{2p}\ln[\chi^2+\rho^2]$, where $p$ (positive integer) is the structure parameter. The corresponding potentials (\ref{potentialnew}) read
    \begin{eqnarray}
    V_{L,R}\!\!&=&\pm\frac{8a\eta}{9}
       \tanh^{2p-2}(ky)        \text{sech}^{4/9a}(ky)
       e^{\frac{2(1-3a)}{9a}\tanh^2(ky)}\nonumber\\
    &\times&\Bigg\{2\tanh^4(ky)\left(1\pm9a\eta\tanh^{2p}(ky)\right)\nonumber\\
    &+&\text{sech}^2(ky)\tanh^2(ky)\Big[-9(a+4ap)\nonumber\\
    &-&2(1+9a) \mathcal{F}(y)+(6a-2)\tanh^2(ky)\nonumber\\
    &\mp&36a\eta \mathcal{F}(y)\tanh^{2p}(ky)\Big]\nonumber\\
    &+& \mathcal{F}(y) \text{sech}^4(ky) \Big[9a(2p-1)+(2-6a)\tanh^2(ky)\nonumber\\
    &\pm&18a\eta \mathcal{F}(y) \tanh^{2 p}(ky)\Big]\Bigg\},
    \label{pot}
    \end{eqnarray}
where $\mathcal{F}(y){\equiv}p\ln\left(\frac{(1-2 a) }{a} \text{sech}^2(k y)\right)$. The asymptotic behaviors of the above effective potentials are described as follows
    \begin{eqnarray}
        V_{L,R}(0) &=&\left\{\begin{array}{cl}
                                  \pm8a\eta\ln(\frac{1-2a}{a}), & ~~~p=1 \\
                                  0, & ~~~p\ge2
                                \end{array} \right.
                                \label{Vasymptoticmultiscalar1} \\
        V_{L,R}(\infty)&\to& 0.\label{Vasymptoticmultiscalar2}
    \end{eqnarray}
In the following discussion, we take positive coupling $\eta$ in order to localize the left-chiral fermion zero mode.
We can see that the parameter $a$ will affect the asymptotic behavior (\ref{Vasymptoticmultiscalar1}) if $p=1$. In order to investigate the property of the potentials at $y=0$, we calculate the first-order and second-order derivatives of $V_{L,R}(0)$ with respect to $a$. We find that when the parameter $a\in(0, 0.412141)$, there is a double-well for $V_{L}(y)$, while for $a\in[0.412141, 0.5)$ we get a single-well (see Fig. \ref{figchangea}). The appearance of the well in $V_{R}(y)$ is decided by the two parameters $a$ and $\eta$.
When $p\geq2$, the effective potentials will vanish at $y=0$ and there are a double-well and a single-well for $V_{L}(y)$ and $V_{R}(y)$, respectively (see Fig. \ref{potentialchangep}). The appearance of the well in $V_{R}(y)$ may result in fermion resonances.

The solution of the zero modes based on (\ref{newzero mode}) are
    \begin{equation}
    f_{L0,R0}\propto \left(\sqrt{\frac{1}{a}-2}~\text{sech}(ky)\right)^{\pm2\eta\tanh^{2p}(ky)}.
    \label{zero modes22}
    \end{equation}
Since $\text{sech}(ky)\to 2 \text{e}^{-k|y|}$ and $\tanh^{2p}(ky)\to1$ as $|y|\to\infty$, the asymptotic behaviors of the zero modes are
    \begin{eqnarray}
    f_{L0,R0}(y)|_{y\to\infty}&\propto&e^{\mp2\eta k|y|}.
    \end{eqnarray}
Therefore, the normalization conditions are
    \begin{eqnarray}
    \int_0^\infty{|f_{L0,R0}(z)|^2dz}&=&\int_0^\infty{|f_{L0,R0}(y)|^2e^{-A}dy}\nonumber\\
    &\to& \int_0^{\infty}e^{(\frac{4}{9}\mp 4{\eta}k)|y|}dy < \infty,
    \label{integral}
    \end{eqnarray}
which are turned out to be $\eta>\frac{1}{18a}$ and $\eta<-\frac{1}{18a}$ for the left- and right-chiral fermion zero modes, respectively. In what follows, we take $\eta>\frac{1}{18a}$ in order to obtain localized left-chiral fermion zero mode on the brane. Plots of the zero modes with different values of $a$ and $p$ are shown in Figs. \ref{figchangea} and \ref{potentialchangep}, respectively. From Fig. \ref{figchangea}, it can be seen that the parameter $a$ affects the localization position of the left-chiral zero mode: when $a$ is small, there is a large potential barrier around the origin of extra dimension and the left-chiral zero mode is localized at two sides; when $a$ becomes large, there is a potential well around the origin and the left-chiral zero mode is localized there.
From Fig. \ref{potentialchangep}, one can see that there are two wells and one well for $p\ge2$ and $p=1$, respectively. But the left-chiral zero mode is always localized around the origin of extra dimension.

\begin{figure}[!htb]
    \subfigure[$\eta=1$]{
    \includegraphics[width=0.22\textwidth]{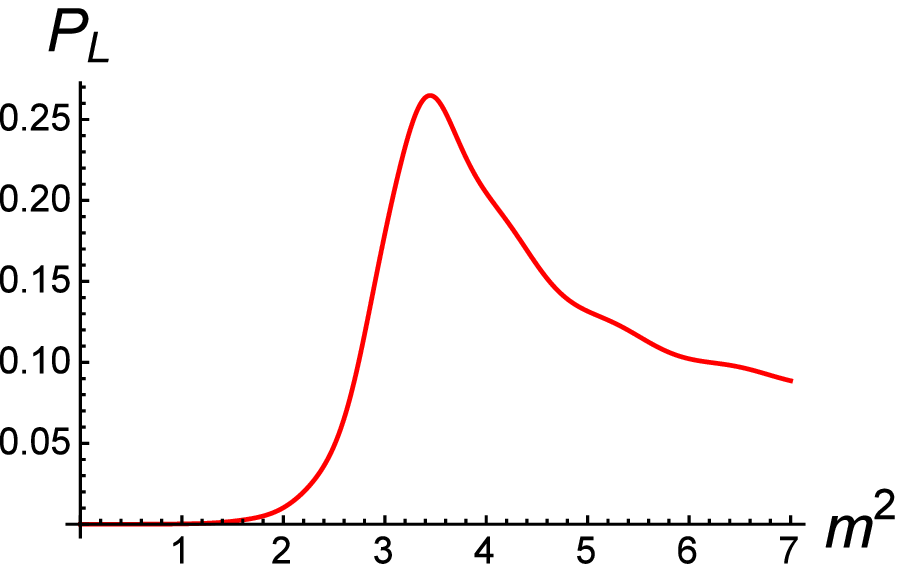}}
    \subfigure[$\eta=1$]{
    \includegraphics[width=0.22\textwidth]{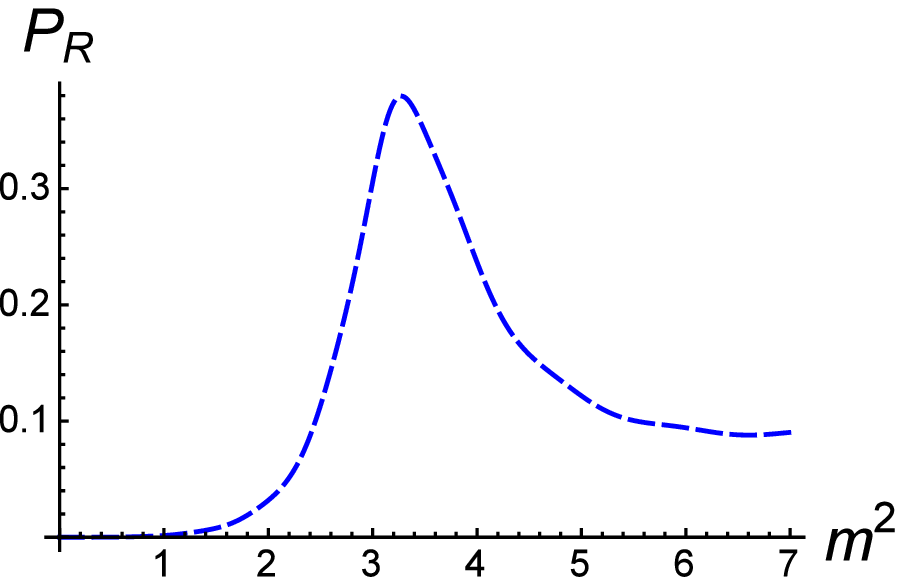}}
    \subfigure[$\eta=3$]{
    \includegraphics[width=0.22\textwidth]{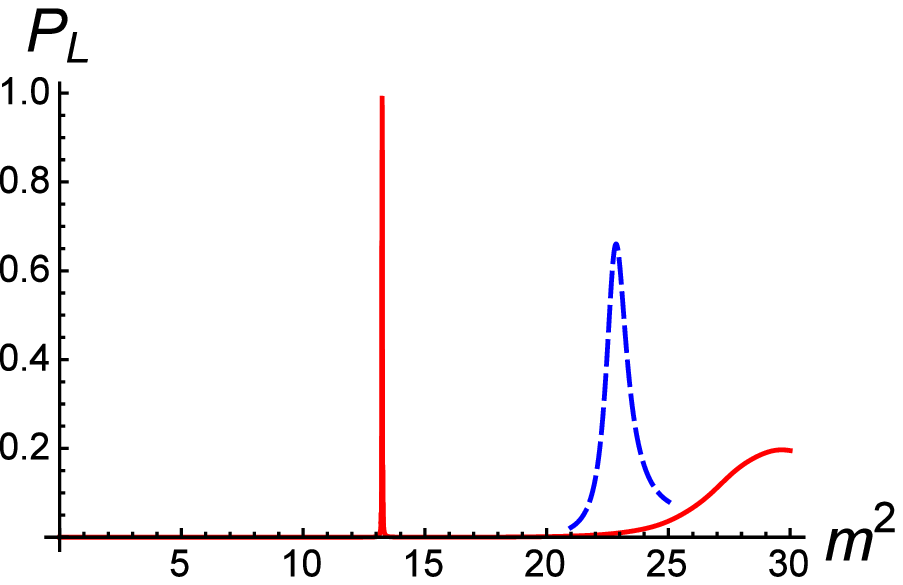}}
    \subfigure[$\eta=3$]{
    \includegraphics[width=0.22\textwidth]{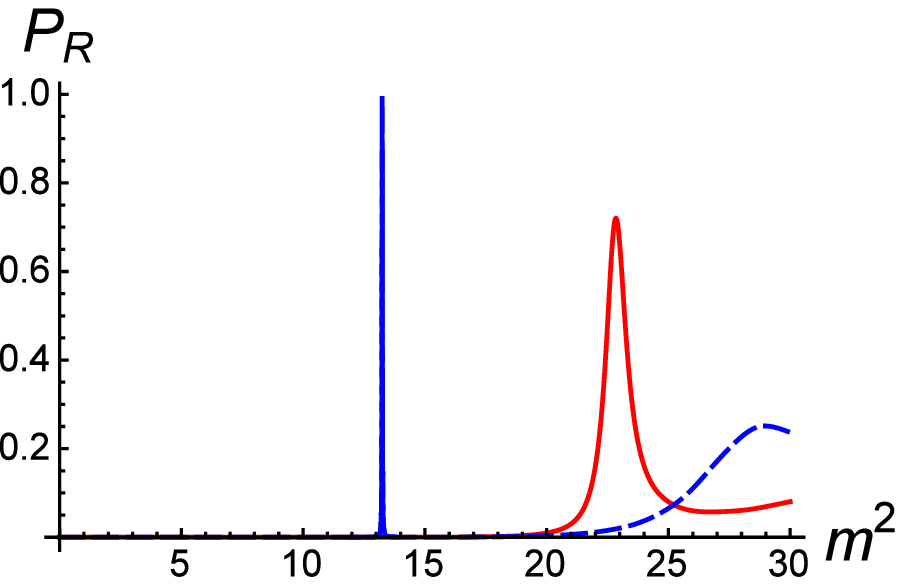}}
    \subfigure[$\eta=5$]{
    \includegraphics[width=0.22\textwidth]{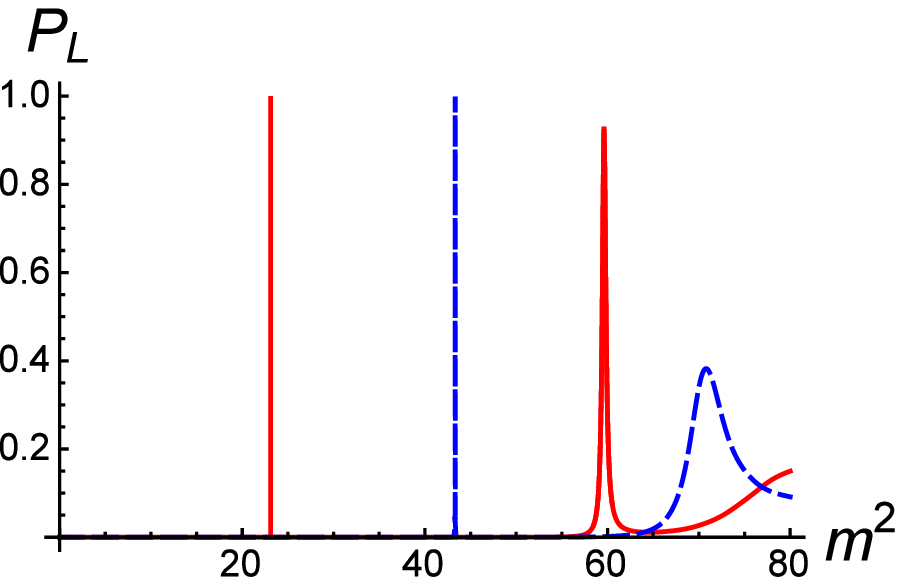}}
    \subfigure[$\eta=5$]{
    \includegraphics[width=0.22\textwidth]{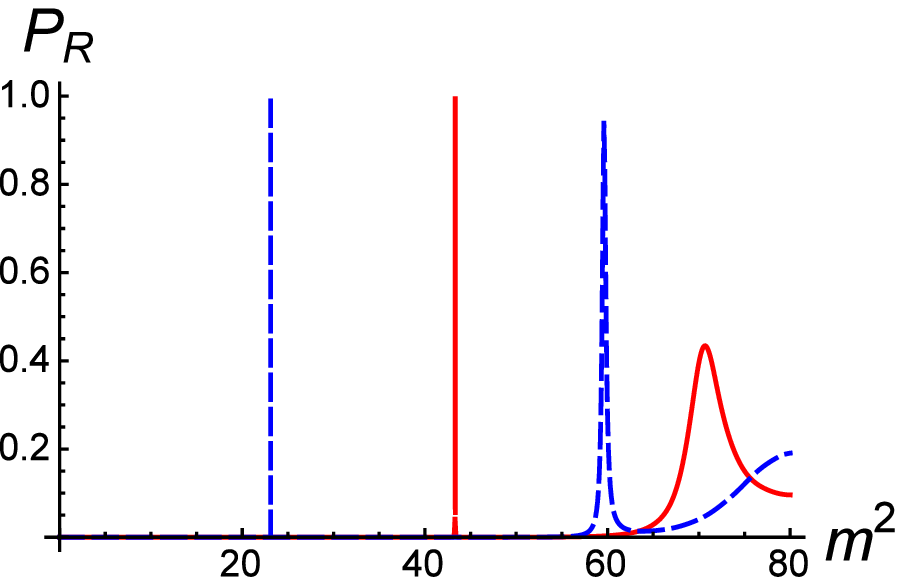}}
    \vskip -4mm \caption{Plots of the probabilities $P_{L,R}$ for the LXCW coupling mechanism. The parameters are set to $a=0.45$ and $p=1$.}
    \label{vlposibility}
    \end{figure}

\begin{figure}[!htb]
    \subfigure[$\eta=1$]{
    \includegraphics[width=0.22\textwidth]{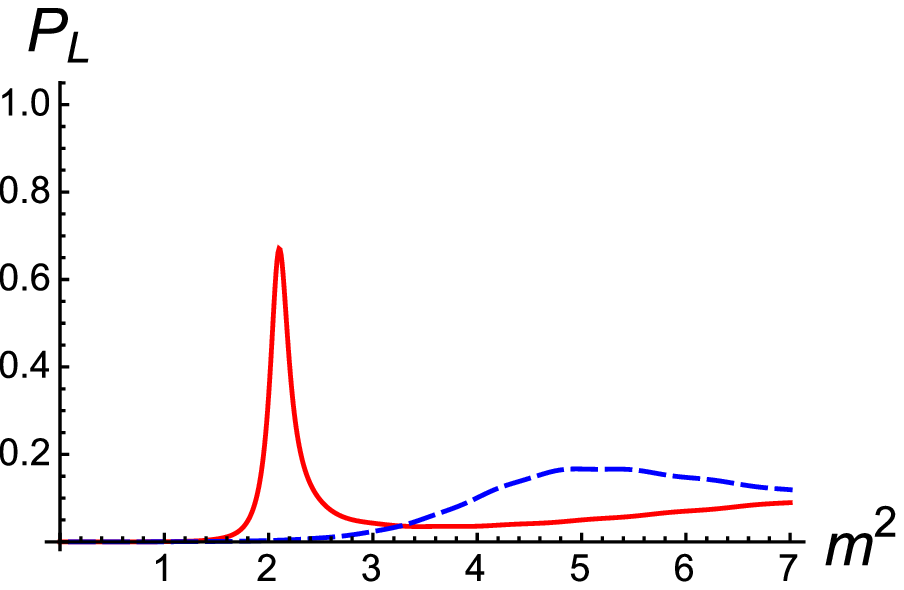}}
    \subfigure[$\eta=1$]{
    \includegraphics[width=0.22\textwidth]{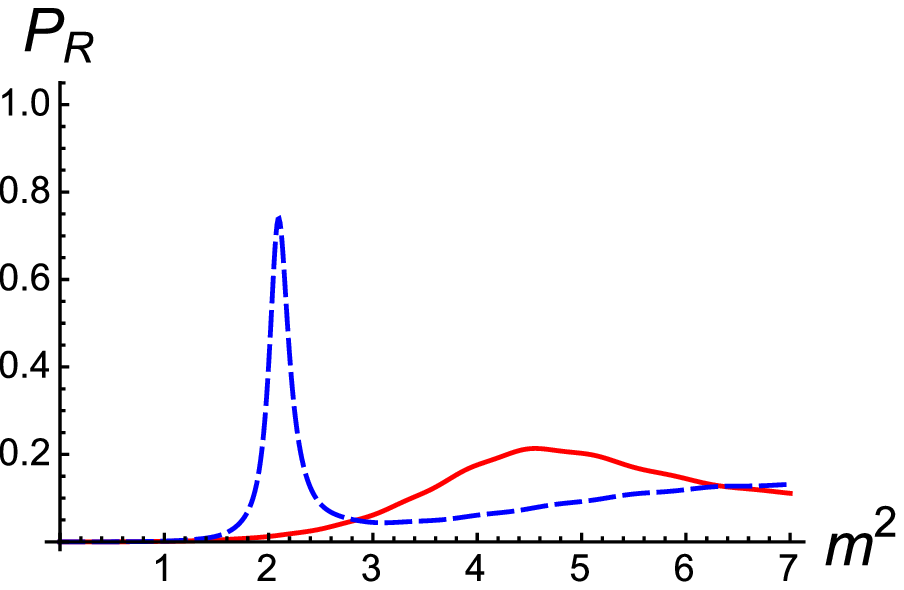}}
    \subfigure[$\eta=3$]{
    \includegraphics[width=0.22\textwidth]{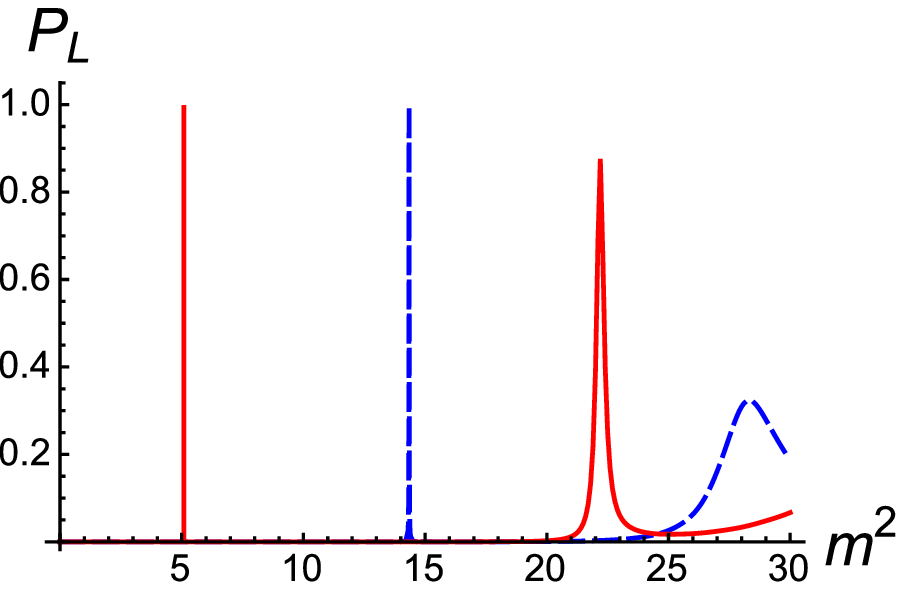}}
    \subfigure[$\eta=3$]{
    \includegraphics[width=0.22\textwidth]{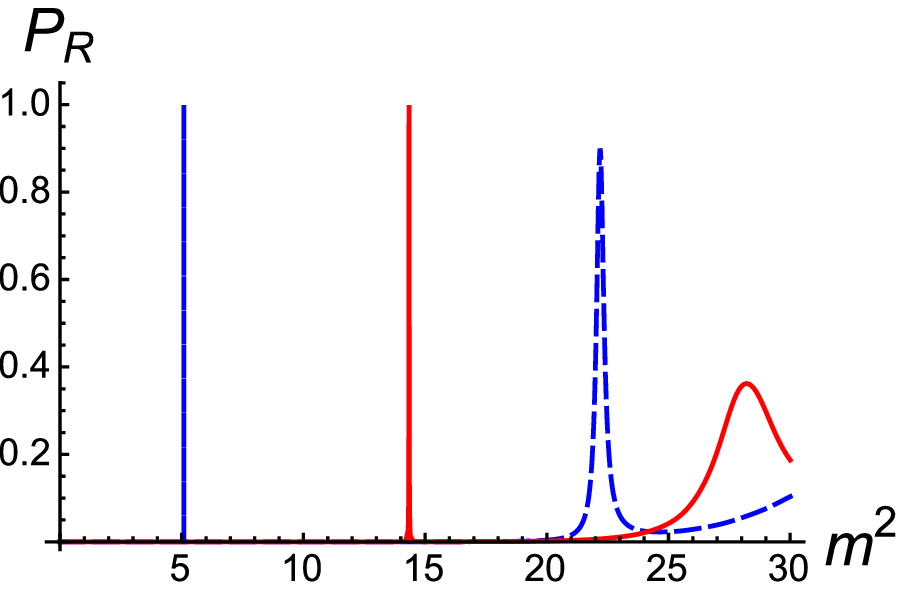}}
    \subfigure[$\eta=5$]{
    \includegraphics[width=0.22\textwidth]{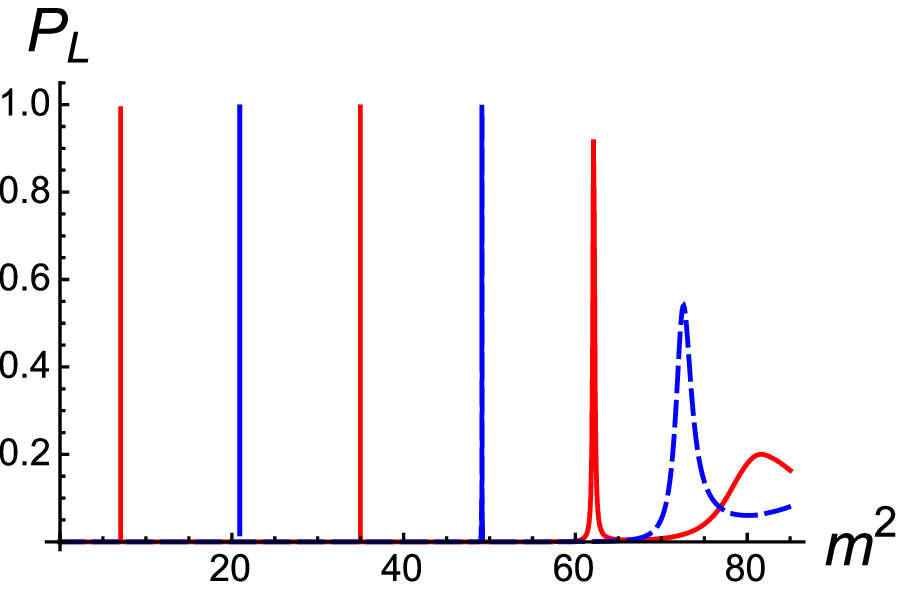}}
    \subfigure[$\eta=5$]{
    \includegraphics[width=0.22\textwidth]{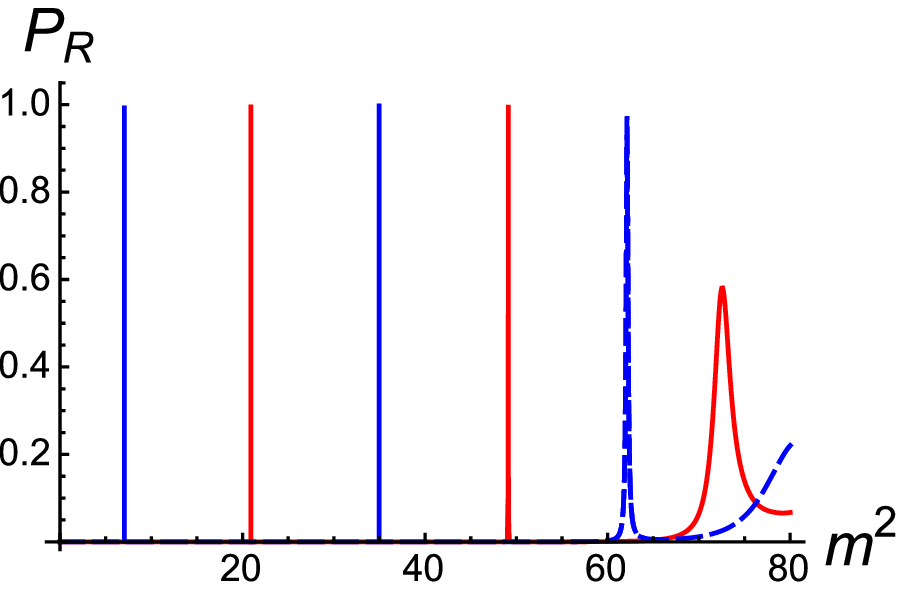}}
    \vskip -4mm \caption{Plots of the probabilities $P_{L,R}$ for the LXCW coupling mechanism. The parameters are set to $a=0.45$ and $p=2$.}
    \label{vlposibilityp2}
    \end{figure}

\begin{figure}[!htb]
    \subfigure[$f_{L1}(z)$]{
    \includegraphics[width=0.22\textwidth]{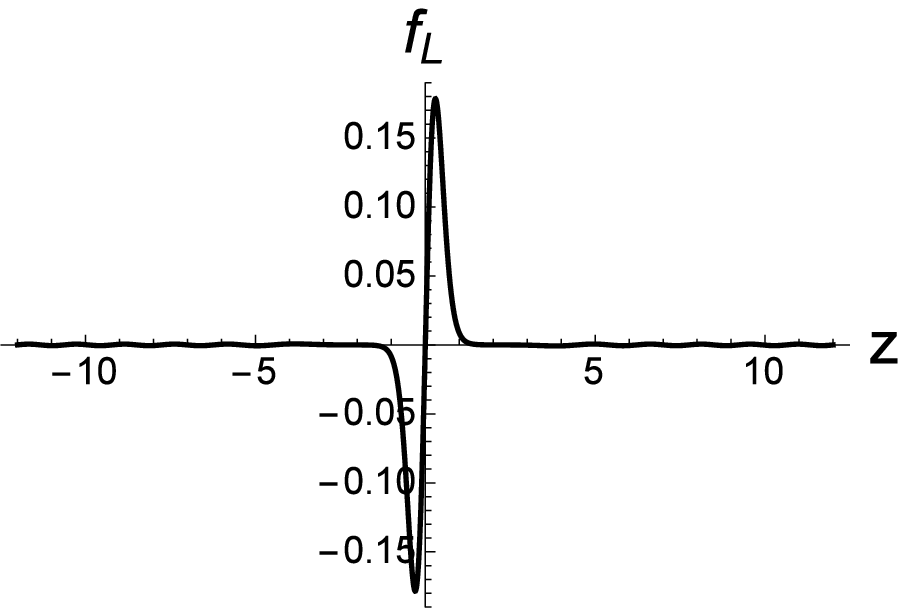}}
    \subfigure[$f_{R1}(z)$]{
    \includegraphics[width=0.22\textwidth]{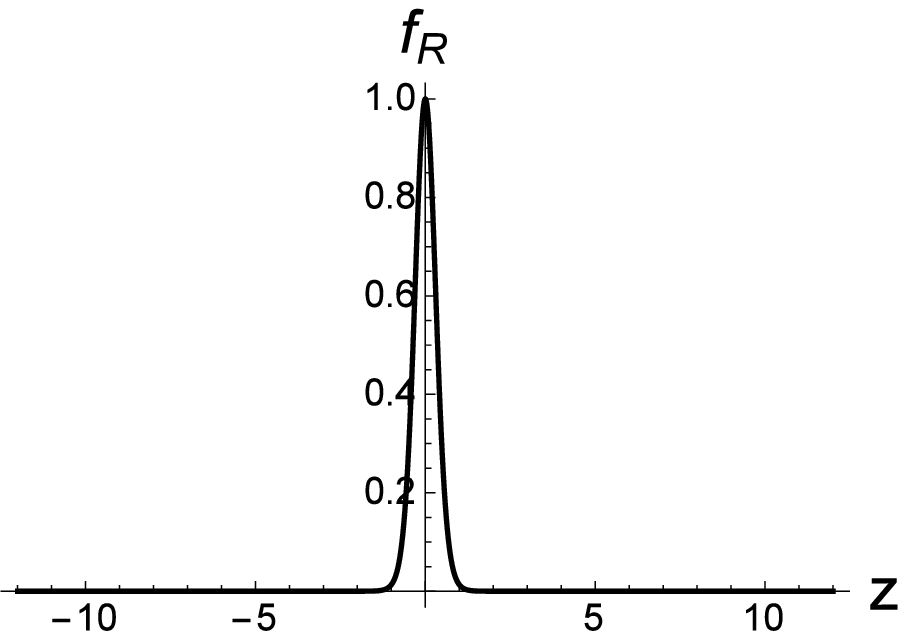}}
    \subfigure[$f_{L2}(z)$]{
    \includegraphics[width=0.22\textwidth]{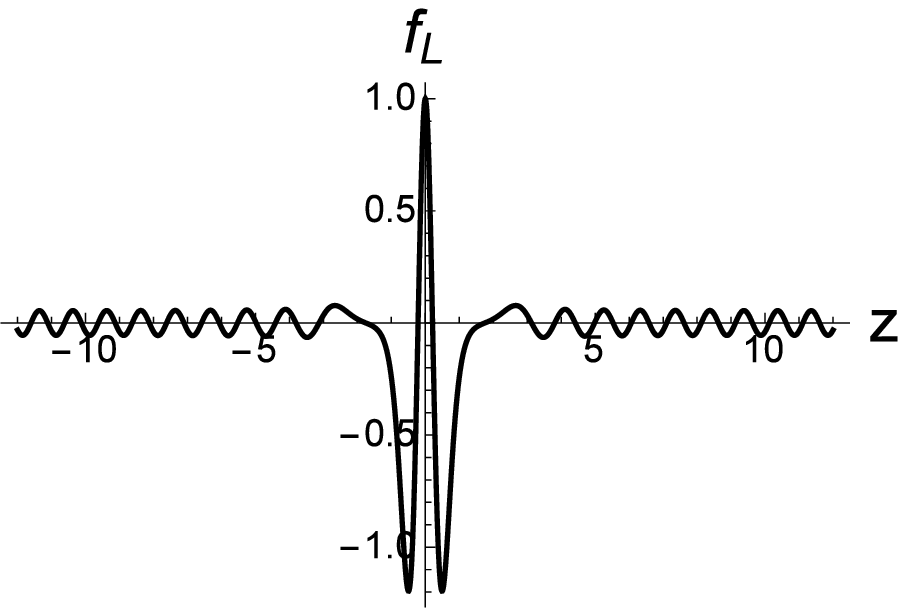}}
    \subfigure[$f_{R2}(z)$]{
    \includegraphics[width=0.22\textwidth]{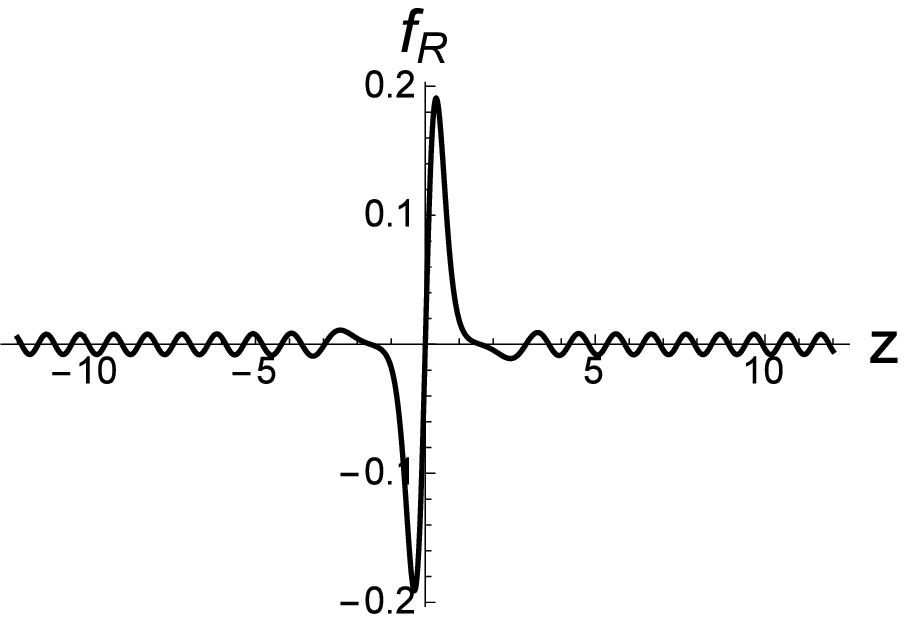}}
    \subfigure[$f_{L3}(z)$]{
    \includegraphics[width=0.22\textwidth]{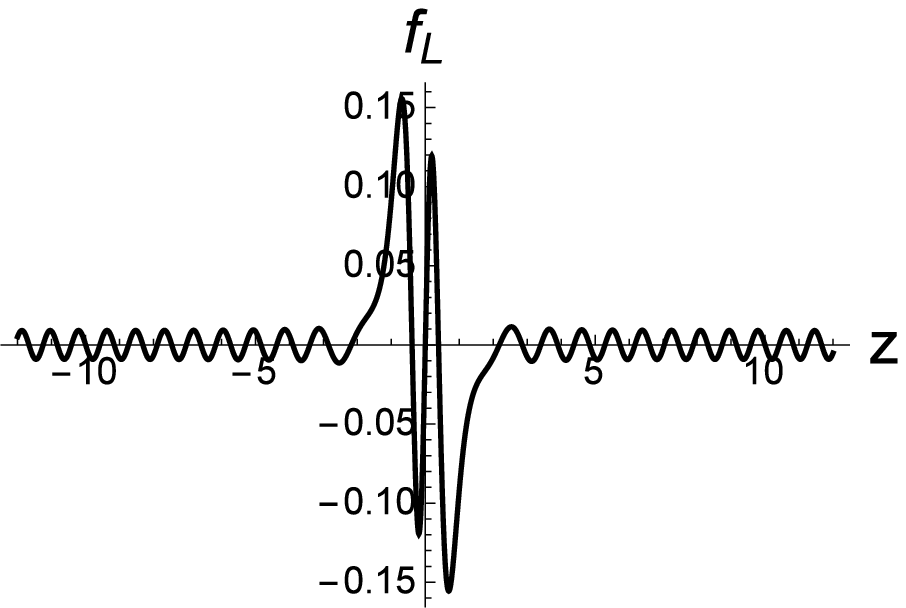}}
    \subfigure[$f_{R3}(z)$]{
    \includegraphics[width=0.22\textwidth]{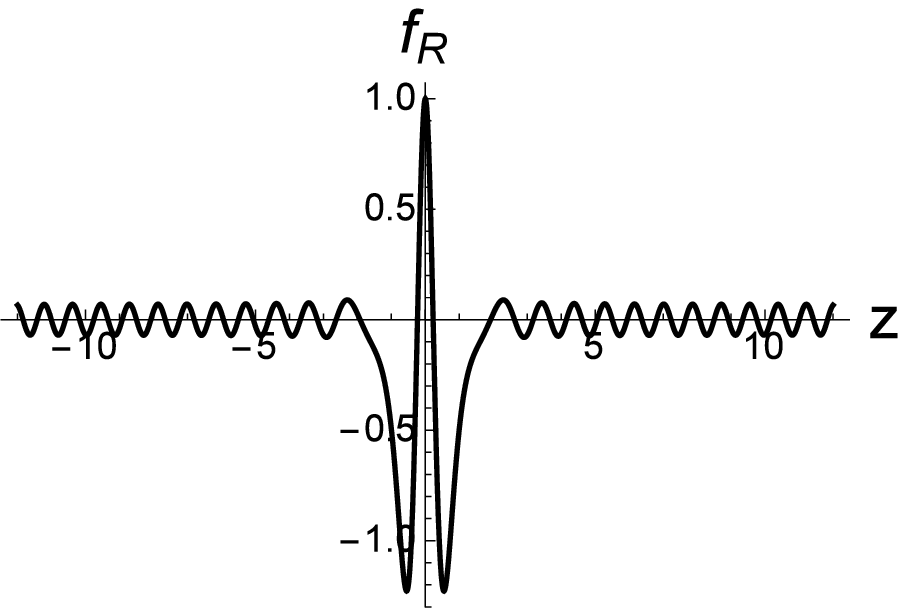}}
    \vskip -4mm \caption{Plots of the resonances of the left- and right-chiral KK fermions for the LXCW coupling mechanism with $\eta=5$ and $p=1$.}
    \label{ryesonances2}
    \end{figure}

Next, we come to massive KK modes. Since the potentials vanish at the boundaries of extra dimension, there is no bound massive KK modes. But we can also use the relative probability to find the fermion resonances. Plots of the relative probabilities $P_{L,R}$ (\ref{zbmin}) are shown in Figs.~\ref{vlposibility} and \ref{vlposibilityp2}, from which we can see that the number of the fermion resonances increasing with the coupling parameters $\eta$ and $p$ since they increase the quasi-potential well of $V_R$. Mass spectrum, width, lifetime, and relative probability of the left- and right-chiral fermion resonances are listed in Table \ref{TableSpectra2} for $a=0.45$, $p=1$, and $\eta=3,~5$. For simplicity we only give the fermion resonant state in Fig. \ref{ryesonances2} with the case of $p=1$ and $\eta=5$.

    \begin{table*}[!htb]
    \begin{center}%\renewcommand\arraystretch{1.4}
    \begin{tabular}{||c|c|c|c|c|c|c|c||}
     \hline
     $\eta$                  & $\text{chirality}$    & $\text{parity}$ & $m^{2}_{n}$  & $m_{n}$    & $\Gamma$              & $\tau$                &  $P$   \\

    \hline \hline

                             &  $\mathcal{L}$ & odd      & 13.2466       & 3.6396    & $2.20\times 10^{-3}$             & $454.9$                & 0.996    \\  \cline{3-8}
                             &                & even     & 22.8513       & 4.7803    & $9.62\times 10^{-2}$             & 10.38                 & 0.661    \\  \cline{2-8}
    \raisebox{2.3ex}[0pt]{3} &  $\mathcal{R}$ & even     & 13.2465       & 3.6396    & $2.10\times 10^{-3}$             & $475.8$                &0.997     \\  \cline{3-8}
                             &                & odd      & 22.8437       & 4.7795    & $9.95\times 10^{-2}$             & 10.05               &0.720     \\  \cline{3-8}

    \hline\hline

                             &                & odd      & 23.1095        & 4.8072   & $6.86\times10^{-7}$        & $1.46\times10^6$    & 0.999    \\  \cline{3-8}
                             & $\mathcal{L}$  & even     & 43.3217        & 6.5819   & $6.08\times10^{-4}$        & $1.65\times10^3$   & 0.998    \\  \cline{3-8}
                             &                & odd      & 59.6198        & 7.7214   & $2.65\times10^{-2}$        & 37.66              & 0.930    \\  \cline{3-8}

    \cline{2-8}
    \raisebox{2.3ex}[0pt]{5} &                & even     & 23.1085        & 4.8071   & $7.90\times10^{-7}$        & $1.27\times10^6$  & 0.999    \\  \cline{3-7} \cline{3-8}
                             & $\mathcal{R}$  & odd      & 43.3215        & 6.5819   & $6.08\times10^{-4}$        & $1.65\times10^3$              & 0.998    \\  \cline{3-8}
                             &                & even     & 59.6165        & 7.7212   & $2.66\times10^{-2}$        & 37.66              & 0.949    \\  \cline{3-8}
       \hline
    \end{tabular}\\
    \caption{Mass spectrum $m^2_n$ and $m_n$, width ($\Gamma$), lifetime ($\tau$), and relative probability ($P$) of the left- and right-chiral KK fermion resonances. The parameters are set to $a=0.45$, $p=1$, and $\eta=3,~5$.}
    \label{TableSpectra2}
    \end{center}
    \end{table*}

In this subsection, we considered localization and resonances of a bulk fermion on a multi-scalar-field thick brane with the LXCW coupling. The conclusion is similar to the case of the single-scalar-field thick brane.
The resonances of fermion based on the Yukawa coupling mechanism had been investigated in Ref.~\cite{Liu2009}. The authors considered the coupling function as $F_2=\phi^q\chi\rho$ with odd $q\ge1$ and found that the left-chiral zero mode can not be localized on the brane because of the lump configurations of $\chi$ and $\rho$. In order to localize the left-chiral zero mode on the brane the authors adopted the coupling $F_2=\phi^q$ and found the localization condition is $\eta>\frac{2}{9}$. In the LXCW coupling mechanism the left-chiral zero mode can be localized on the brane with the lump configurations of $\chi$ and $\rho$ and the corresponding localization condition is $\eta>\frac{1}{18a}$. For the Yukawa coupling mechanism the parameter $a$ in the brane solution does not affect the localization and localization position of the left-chiral zero mode \cite{Liu2009}, but for the LXCW coupling mechanism it does (see Fig. \ref{figchangea}).

\begin{figure}[!htb]
    \subfigure[Yukawa coupling]{
    \includegraphics[width=0.22\textwidth]{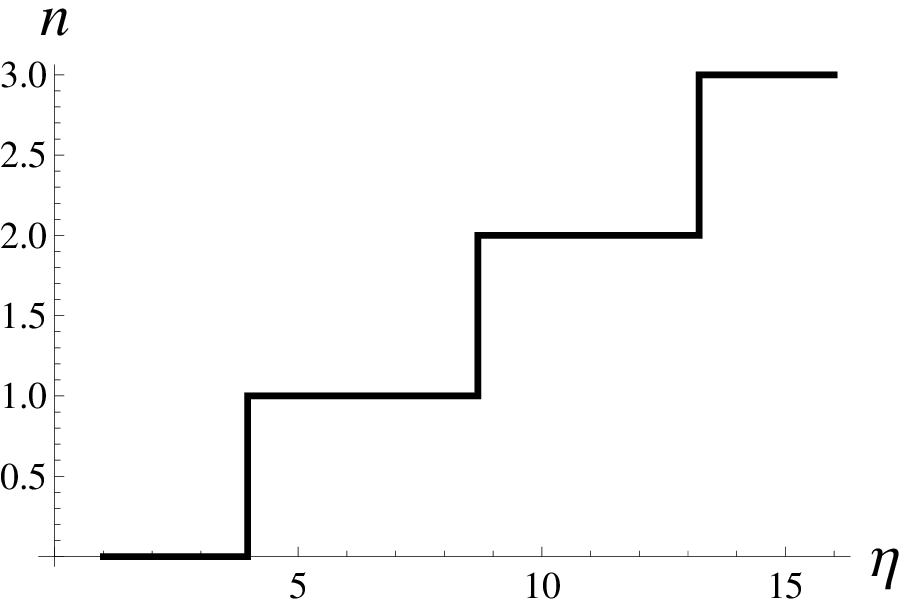}}
        \subfigure[LXCW coupling]{
    \includegraphics[width=0.22\textwidth]{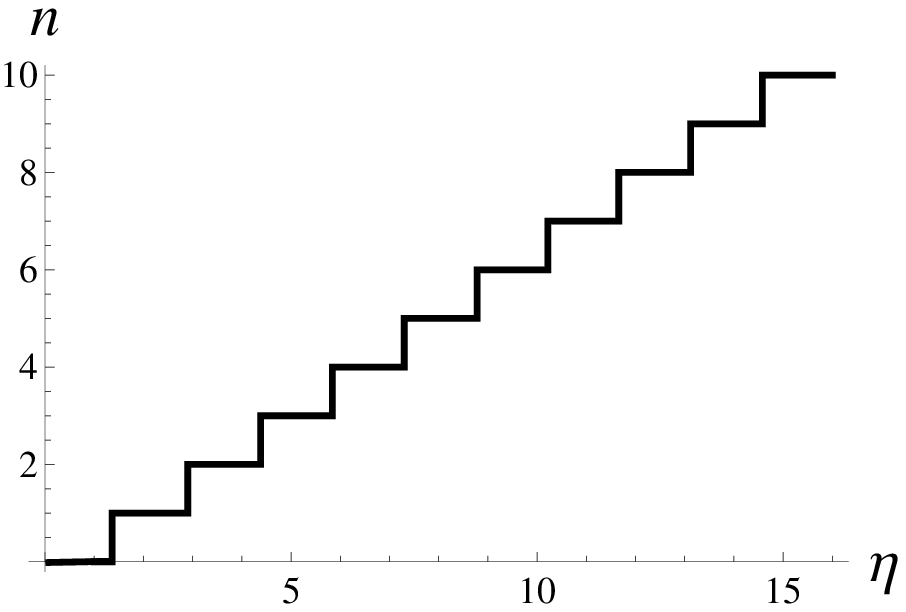}}
    \vskip -4mm \caption{Plots of the relationship between the number $n$ of the resonances and the coupling $\eta$ for the single-scalar-field model. (a) Yukawa coupling $\eta\bar{\Psi}\text{arcsinh}^{2q-1}(b\phi)\Psi$ with $q=1$. (b) LXCW coupling $\eta\bar{\Psi}\Gamma^M\partial_M \text{arcsinh}^{2p}(b\phi) \gamma_5\Psi$ with $p=1$.}
    \label{ResonancesNumber1}
    \end{figure}

\begin{figure}[!htb]
    \subfigure[Yukawa coupling]{
    \includegraphics[width=0.22\textwidth]{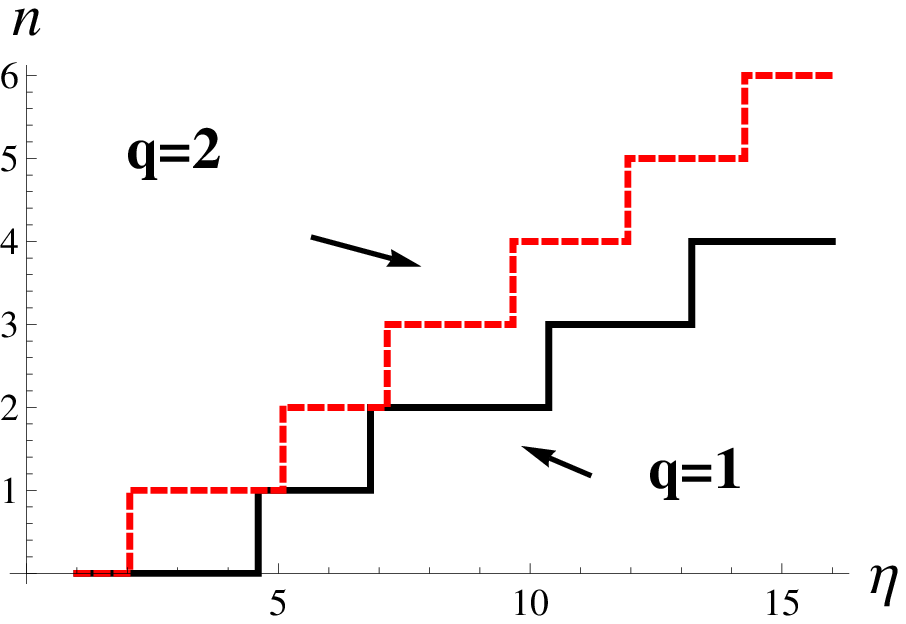}}
    \subfigure[LXCW coupling]{
    \includegraphics[width=0.22\textwidth]{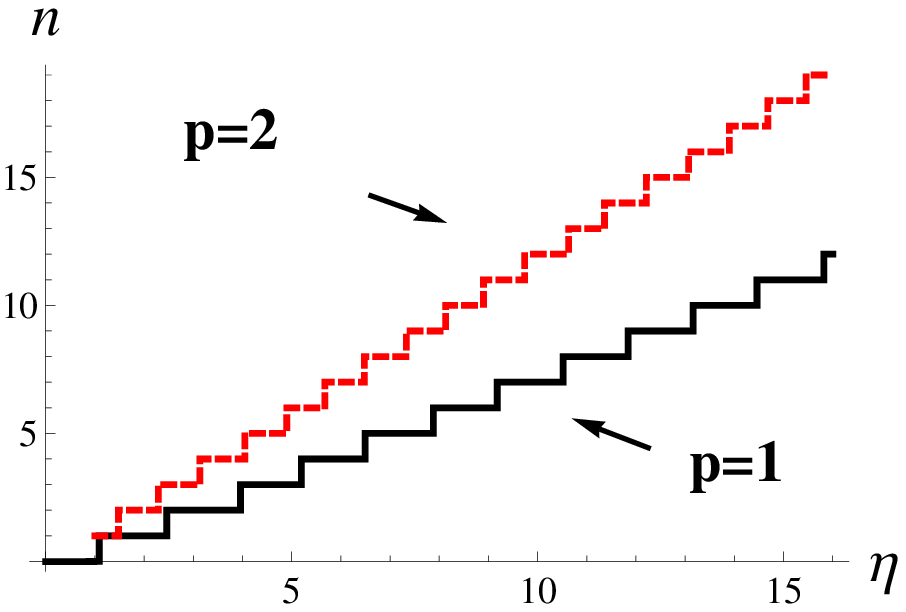}}
    \vskip -4mm \caption{Plots of the relationship between the number $n$ of the resonances and the coupling $\eta$ for the multi-scalar-field model. (a) Yukawa coupling $\eta\bar{\Psi}\phi^{2q-1}\Psi$ with $a=0.45,~q=1,~2$. (b) LXCW coupling $\eta\bar{\Psi}\Gamma^M\partial_M (\phi^{2p}\ln[\chi^2+\rho^2])\gamma_5\Psi$  with $a=0.45,~p=1,2$.}
    \label{ResonancesNumber2}
    \end{figure}

The relationships between the number $n$ of the resonances and the coupling $\eta$ for the two models considered in the paper are respectively shown in Figs. \ref{ResonancesNumber1} and \ref{ResonancesNumber2}, which show a simple linear relationship between them, i.e., the number of the resonances will increase linearly with $\eta$ for both Yukawa coupling and LXCW coupling. The number $n$ also increases with the parameter $p$ or $q$. In both models with our chosen coupling functions and $p=q=1$, LXCW coupling will result in more fermion resonances.

\section{Conclusion}\label{Conclusion}

In this paper, we first reviewed the localization mechanisms for fermions and the numerical method to find fermion resonances.  Then we investigated localization and resonances of a bulk fermion on the single-scalar-field and multi-scalar-field thick branes based on the Yukawa and LXCW coupling mechanisms.

For the single-scalar-field model, we considered respectively the Yukawa coupling $\eta\bar{\Psi}F_2(\phi)\Psi$ and LXCW coupling $\eta\bar{\Psi}\Gamma^M\partial_MF_1(\phi)\gamma_5\Psi$, where $F_2=\lambda_1 \phi + \lambda_2\sinh(b\phi) + \lambda_3 \text{arcsinh}^{2q-1}(b\phi)$ and $F_1=\text{arcsinh}^{2p}(b\phi)$ with $q$ and $p$ integers. For the Yukawa coupling mechanism we mainly focused on the resonances of fermion because localization of a fermion had been investigated in Ref.~\cite{Liang2009}, so we only consider $F_2=\text{arcsinh}^{2q-1}(b\phi)$ for simplicity.
For the multi-field thick brane model we only considered the LXCW coupling mechanism and chose $F_1=\phi^{2p}\ln[\chi^2+\rho^2]$, with which the left-chiral fermion zero mode can be localized on the brane under the condition $\eta>\frac{1}{18a}$.

The structures of the effective potentials of the left- and right-chiral fermion KK modes are determined by the parameters $q$ or $p$ and coupling constant $\eta$ (the depth and width of the effective potentials increase with them). The effective potential $V_L$ of the left-chiral fermion KK modes is volcano-like when $p=1$ or $q=1$ (we did not consider the affects of $a$ and other parameters). When $q\ge2$ or $p\ge2$, the effective potential $V_{L}$ has a double-well, and interestingly $V_{R}$ has a quasi-well. We obtained the massive resonant fermions on the flat branes. It was found that the double-well potentials will product more fermion resonances than the single-well ones and there is a simple linear relationship between the number of fermion resonances and the coupling parameter $\eta$. Although the effective potentials have different structures, the mass spectra and lifetimes of left- and right-chiral fermion resonances are the same in both coupling mechanisms. The reason is that $H_L=U^{\dag}U$ and $H_R=UU^{\dag}$ in (\ref{operator2}) are conjugated supersymmetric partner Hamiltonians  with the superpartner potentials $V_L$ and $V_R$. Hence a massive Dirac fermion with finite lifetime consist of a pair of left- and right-chiral fermion resonant KK modes.

%\section*{Acknowledgement}
\acknowledgments{
This work was supported in part by the National Natural Science Foundation of China (Grants No. 11375075  and No. 11522541)}, and the Fundamental Research Funds for the Central Universities (Grant No. lzujbky-2015-jl1).

%%%%%%%%%%%%%%%%%%%%%%%%%%%%%%%%%%%%%%%%%%%%%%%%%%%%%%%%%%%%%%%%%%%%%%%%%%%%%%%%%%%%%%%

\end{document}